\documentclass[authoryear]{elsarticle}
\usepackage{lineno,hyperref}
\modulolinenumbers[1]
\pdfoutput=1 
\graphicspath{ {/} }
\usepackage{floatrow}
\usepackage{rotating}
\usepackage[margin=2.5cm]{geometry}
\journal{}

\begin{document} 
\begin{frontmatter}

\title{Stratigraphic and sedimentologic framework for tephras in the Wilson Creek Formation, Mono Basin, California}
\author[geo_ub]{Qingyuan Yang \corref{cor1}}
\ead{qyang5@buffalo.edu}

\author[geo_ub]{Marcus Bursik}
\author[geo_ub]{Sol\`ene Pouget{\textdagger}}

\address[geo_ub]{Department of Geology, University at Buffalo, SUNY, Buffalo, NY 14260, United States}
\cortext[cor1]{Corresponding author}

\begin{abstract}
	Numerous tephra layers occur within the late Pleistocene Wilson Creek Formation, where they are interbedded with lacustrine deposits of Lake Russell, the ancestor of present-day Mono Lake. Most of the tephra layers are rhyolitic in composition, and were produced from the Mono Craters. We present detailed stratigraphy and sedimentology of the tephra layers, sampled at twelve outcrops near the shoreline of Mono Lake and the Mono Craters, and implement grain size, componentry, and surface morphology analysis to characterize their physical properties. Sub-unit correlation is proposed for certain tephra units. Noticeable features of the tephras, such as the occurrence of low-density rounded or highly vesicular pumice within certain sub-units, are highlighted. The abundant obsidian, lithics, and ostracods within many sub-units suggests that the associated eruption pulses involved water-magma interaction. Eruptions from the Mono Craters during the late Pleistocene were consistent neither in frequency nor volume. The Mono Craters were most active during the eruption of Sequences C and A, and Pleistocene volcanic activity reached its peak during the eruption of C11. Detailed interpretation of tephra layer B7 and tephras in Sequence A is given, which includes the number of eruption pulses, vent location, dispersal direction, occurrence of pyroclastic flow and surge, and other features and processes. Tephras in Sequence B may have had their vents located in the southern half of the Mono Craters, or are smaller in volume (except for B7), compared to the other tephras in the Wilson Creek Formation. Vents for A4, A3, and A1 are located near the northern end of the Mono Craters. The stratigraphy of B7 and A1 suggests an unstable depositional environment, which can be used to help constrain the water-level history of Lake Russell during the late Pleistocene.
\end{abstract}


\end{frontmatter}

\bibliographystyle{elsarticle-num-names}

\section{Introduction}
The Wilson Creek Formation (WCF) surrounds Mono Lake in eastern central California \citep{M14}, and is mainly composed of lacustrine deposits formed in Pleistocene Lake Russell \citep{L68, vazquez2012high}, the ancestor of present-day Mono Lake.  Numerous tephra fall beds from explosive volcanic eruptions are preserved within the WCF \citep{L68}. Most tephras within the WCF are rhyolitic in composition, and are thought to be originated from the Mono Craters \citep{L68}.  Tephras in the WCF are divided into five sequences (A-E), A1 (top) - E20 (bottom), and have been used to establish stratigraphic framework and chronology for major paleomagnetic and climate change records of the area (e.g., \citealp{Zimmerman06,Cox2012,vazquez2012high}). Their geochemical compositions have been used to identify the corresponding vent \citep{M14}, and to study the magmatic system beneath the Mono Craters \citep{L68,Bur1989,M14}. Our understanding of these deposits has been continuously updated from fieldwork and laboratory analysis (e.g.,  \citealp{M14}). However, physical properties of the tephras have been rarely investigated \citep{L68}, but are of great importance to the reconstruction of the volcanic history of the Mono Craters during the late Pleistocene. Basic but key information about the deposits, such as total volume and possible vent location, is still unclear. From a hazard management perspective, the active volcanism of the region (e.g., \citealp{miller1985holocene,sieh1986most,hildreth04,bursik2014deposits}) poses potential threats to the local community, infrastructure, and aviation. Studying these older deposits can greatly reduce the uncertainty in the assessment of potential future hazards \citep{bevilacqua2017bayesian,bevilacqua2018late}.

In the present work, we provide a detailed characterization of the stratigraphy and physical properties of the tephra deposits in the WCF, based on outcrops exposed near the shoreline of Mono Lake and the Mono Craters. With these data, we are able to correlate between sub-units from one location to another, highlight noticeable features, and make interpretations about the deposits and their corresponding eruptions. Our work shows that each of the eruptions that produced the WCF tephras are composed of different pulses. The eruptions could have migrated vents. Their eruptive products display a variety of characteristics, and have distinct dispersal patterns. Our work provides qualitative and quantitative constraints to some of these eruptions. We address the uncertainty in our discussion by proposing different possible scenarios, which can be used as working hypotheses for future study. The present work summarizes physical properties of the tephras in the WCF at a number of sample sites within the Mono Basin, and represents a necessary step towards a comprehensive understanding of the tephras in the WCF and their corresponding eruptions. The tephra stratigraphy and sub-unit correlation could act as a reference for and benefit future studies on the WCF tephras at more distal sites.

\section{Geological background}
The volcanic activity in the Mono Lake-Long Valley region of eastern central California has persisted for about 4 Ma, and is recognized as a framework for six successive (spatially discrete) foci of silicic magmatism \citep{hildreth1986ring,hildreth04}. The initial activity includes an episode of precaldera basalt-trachyandesite-dacite-rhyolite volcanism (4.0-0.8 Ma; \citealp{hildreth1986ring,bailey2004}). The eruption of the Bishop Tuff at 760 ka released $\sim$600 km$^3$ of high-silica rhyolite from a subterranean chamber whose subsidence produced the Long Valley caldera \citep{bailey1976volcanism,hildreth1986ring,hildreth04}. Postcaldera events include many eruptions of rhyolitic magma within the western half of the caldera \citep{bailey1976volcanism,hildreth04}.

The most recent volcanism centers around Mammoth Mountain and the Mono-Inyo craters chain.  Near the southwestern rim of Long Valley caldera, a series of dome building eruptions between $\sim$120 ka and 58 ka formed the dacitic to rhyodacitic Mammoth Mountain lava dome complex \citep{mahood2010new}. The Mono-Inyo Craters volcanic chain, extending 25 km north of Mammoth Mountain to Mono Lake, was produced during the late Pleistocene to Holocene, and is chemically distinct from both the Long Valley and Mammoth Mountain magmatic systems \citep{L68,miller1985holocene,kelleher1986mono,sieh1986most,bailey2004,hildreth04,bursik2013digital}. The Mono-Inyo chain is one of the youngest areas of rhyolitic volcanism in the western United States \citep{M14}, with the latest large eruptions taking place at  $\sim600$ years ago at the northern and southern ends of the Mono-Inyo Craters \citep{miller1985holocene, sieh1986most,hildreth04}.

\begin{figure}[H]
	{\includegraphics[width=\textwidth]{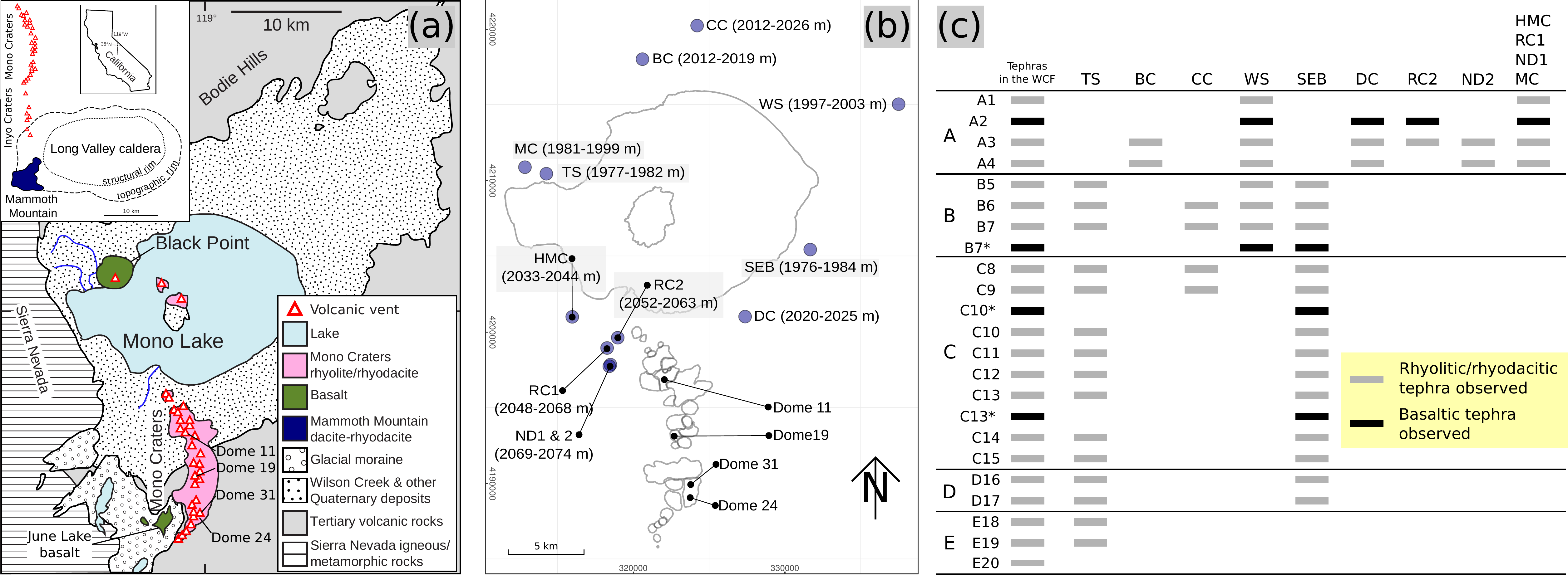}}
	\caption{a: General geologic map of the Mono Basin showing the distribution of the Wilson Creek Formation and other Quaternary deposits. This figure takes reference, and is modified from \cite{M14}. Top left inset figure shows the location of the Mono Craters with respect to the Long Valley caldera, Inyo Craters, and Mammoth Mountain.  b: Sample sites with respect to the Mono Craters and Mono Lake. The elevation range of the observed tephras at each sample site is listed in bracket. The range is given based on the elevation extracted from a local 30-by-30 meter resolution Digital Elevation Model at the highest point near the sample site (within $\sim$120 m from the sample site; upper limit) and at the base of the closest stream cut or cliff (lower limit; no digging was done to sample the tephras) c: Tephra units observed at each site are marked.}
	\label{ggb}
\end{figure}

The Mono Craters are composed of at least 28 overlapping lava domes, flows and craters of high-silica rhyolite, and one dacitic dome near the northern end of the chain (\citealp{L68,marcaida2015resolving}; Fig. \ref{ggb}a and b).  Studies \citep{sieh1986most,Bur1989} on the most recent eruptions from the Mono Craters suggest that an average volume of a single dome can be estimated and used to estimate the number of dikes for the Mono Craters. Based on this, it is estimated that the Mono Craters were fed by approximately 30 dikes \citep{Bur1989}. A recent geochemical study further divides Dome 24 into two separate domes (\citealp{marcaida2015resolving}, Fig. \ref{ggb}a and b).

\begin{figure}[H]
	{\includegraphics[width=\textwidth]{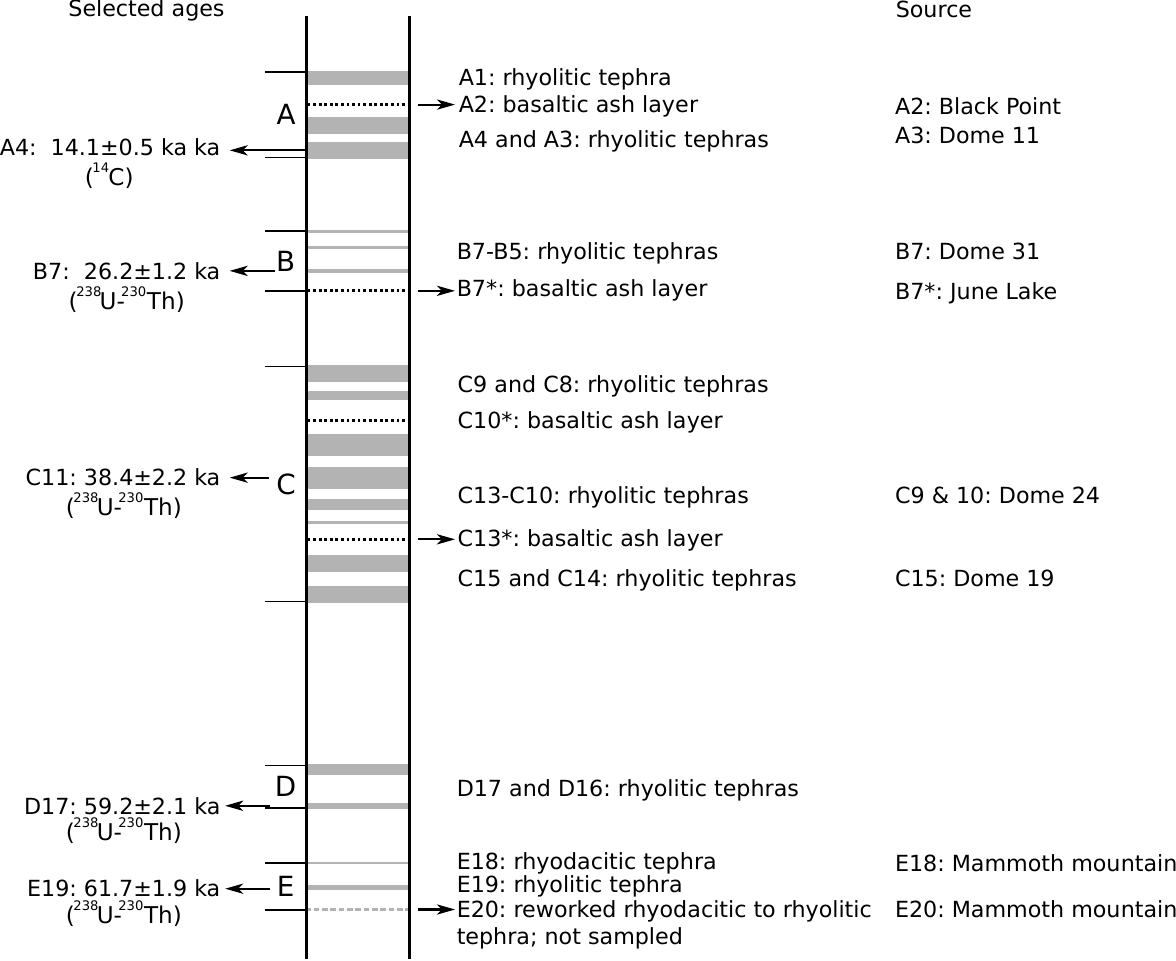}}
	\caption{Brief composite stratigraphy of tephra layers in the WCF. Stratigraphic position of each tephra unit is not scaled. Ages of E19, D17, C11 take reference from \cite{vazquez2012high}, B7 from \cite{marcaida2015resolving}, and A4 from \cite{benson1998correlation}. Marked tephra sources take references from \cite{L68,bursik1993late,M14,marcaida2015resolving}. See text for more details.}
	\label{composite_strat}
\end{figure}

There are at least 23 tephra layers preserved within the WCF (Figs. \ref{ggb}c and \ref{composite_strat}). Their ages range from $\sim$62 ka to 12.92 ka \citep{benson1998correlation,Zimmerman06,vazquez2012high}, and are mostly rhyolitic in composition \citep{L68}.

 They are divided into five sequences (oldest to youngest: Sequences E-A), and are numbered from E20 (oldest) to A1 (youngest) by \cite{L68}. In the present work, we refer to a layer by their corresponding sequence plus the layer number. The other three tephra layers are thin and basaltic  in composition (C13*, C10*, and B7*). They are named based on the stratigraphically closest rhyolitic ash, and are marked with asterisk to denote the difference by \cite{L68}. These notations have been consistently used in previous studies \citep{L68,zimmerman2011high,M14}, and are thus adopted here. Non-rhyolitic ashes also include the basaltic bed A2, rhyodacitic bed E18, and bed E20 characterized by heterogeneous chemistry (rhyodacitic to rhyolitic; minor reworking; see \citealp{M14} for more details on the geochemical properties of ash layer E20). Rhyolitic tephras in the WCF and domes of the Mono Craters are correlated one to another based on their compositional similarities, and E18 and E20 are correlated with Mammoth Mountain based on their similar silica content and titanomagnetite composition \citep{M14}.  Tentative correlations between C15 and Dome 19, C9 and C10 and Dome 24, B7 and Dome 31, A3 and Dome 11 (Fig. \ref{ggb}a and b) have been proposed \citep{M14,marcaida2015thesis,marcaida2015resolving}.  Basaltic ash B7* was  correlated to the June Lake basalt (see \citealp{bursik1993late}, and the map within). A2 was erupted from Black Point Volcano on the northwest shoreline of Mono Lake (\citep{L68}; Fig. \ref{ggb}a). It has been suggested that most domes that are associated with the WCF tephras are now buried \citep{Bur1989}.

Tephras in the WCF were first studied and analyzed by \cite{L68} at several outcrops made by stream and wave cut, and have been investigated by numerous workers because of their importance to lake level and paleoclimate \citep{L68,chen1996edge,benson1998correlation,zimmerman2011high,zimmerman2011freshwater,vazquez2012high}.  Nevertheless, the stratigraphy within each tephra unit, correlation between sub-units at different sample sites, and their eruptive history have not been investigated. Little in fact is known about these tephra units from a physical volcanological perspective, which motivates the current work.

\section{Sample sites and methods}
Along the shoreline of Mono Lake and in the vicinity of the Mono Craters, we have found twelve outcrops with tephra preserved within the WCF (Fig. \ref{ggb}b and c), namely ``north of dump'' (Mono County dump, off State Highway 120; ND1 and ND2), Rush Creek (RC1 and RC2), Horse Meadow Creek (HMC), Mill Creek (MC), type section (TS), Bridgeport Creek (BC), Cottonwood Creek (CC), Warm Spring (WS), Southeast Bluff (SEB), and Dry Creek (DC). ND1 and ND2, RC1 and RC2, and HMC are proximal to the Mono Craters, and are thus called proximal sites in the text. Only tephras in Sequence A are found to be exposed at these locations, and older tephras were not preserved there \citep{L68}. Tephra units observed at each location are marked in Fig. \ref{ggb}c. Tephra units can be directly recognized at MC, TS, and SEB, where the number of tephra units, the distinct features within them (e.g., presence of basaltic layers), the geochemistry, the ages and their relative positions are well-known, and it is not necessary to implement compositional analysis to recognize the tephra units at these sites \citep{L68,chen1996edge,benson1998correlation,benson2003age,zimmerman2011high,vazquez2012high,M14}. It is known that HMC, RC, ND, and DC preserve only Sequence A (A4-A1) tephras \citep{L68}, and the presence of A2, distinct dark scoria layers, further helps us identify these tephra units. Tephra units at WS are recognized based on similar arguments (the presence of dark scoria layers A2 and B7*). Tephra units at BC (A4 and A3) and CC (C9, C8, and B7) are identified solely based on their stratigraphic features (e.g., color, grading, and the number of sub-units), however, they are highly consistent with the corresponding tephra units observed at nearby sites.

Stratigraphy of the WCF tephras is recorded during field work. Samples were collected for grain size, componentry, and surface morphology (mostly fine glass shard and pumice) analyses. They were dried, sieved, and weighed based on grain size (-4 to 4 $\phi$). Componentry of sampled tephra grains coarser than 1$\phi$ was characterized under the microscope with mass weighed and recorded. Finer samples ($> 4\phi$) were analyzed using a Hitachi S4000 Field Emission Scanning Electron Microscope (SEM) with IXRF Energy-dispersive X-ray Spectrometer. Samples were prepared by spreading a small quantity on SEM tape.  We took ten pictures of each SEM tape at different scales. Each picture was then analyzed using the image processing programme ImageJ \citep{rasband1997imagej,schneider2012nih}. Contours of particles were obtained by adjusting the threshold on the original SEM images. Particles that were overlapping in SEM images were separated manually to avoid overestimation of grain size. In practice, as the fine grains were evenly distributed on the SEM tape, this procedure was not always required. A built-in function within ImageJ  was used to measure the major and minor axes as well as the projected area and diameter of each grain. The method proposed by \cite{mills2010shape} was used to calculate the volume of each grain from the 2D images. These measurements were further used as input to GSSTAT with inclusive graphic statistics methods \citep{poppe2004visual} to estimate the grain size distribution of sampled tephras.

\section{Stratigraphic characteristics}
A detailed characterization of the WCF tephras as well as their underlying and overlying sediments is presented in this section. The stratigraphy of each tephra unit reported herein is described to a different level of detail, as features and number of sub-units for each tephra layer could vary significantly from one place to another. The terminology used to describe the grain size of different deposits is defined in Table \ref{term} to avoid confusion.

Since most tephras in Sequences E-B (Figs. \ref{d_overall}-\ref{b_overall_notext}) are observed at two to three sample sites, they are described by sequence for simplicity. For tephras in Sequences E-B, we describe only the units that display distinct and complex characteristics in the text. Features of the rest of the tephra units in Sequences E-B at different sample sites are listed in Tables \ref{ts_e2c_feature}-\ref{seb_b_feature}. Each tephra unit in Sequence A (Figs. \ref{a4p}-\ref{a1p}) is characterized separately due to the complex stratigraphy exposed at proximal sample sites. Contacts between units and sub-units are assumed to be planar and sharp in the text and tables unless otherwise specified.

\subsection{Sequences E and D (E19-D16)}
Tephras in Sequence E were sampled and described only at site TS (northwest of Mono Lake).  Although \citet{marcaida2015resolving} described a new, undeformed ash, E20, below E19, it was found that the heterogeneous chemistry of the layer precluded its being primary.  E19 is thus the lowermost primary tephra sampled (Fig.\ref{d_overall}). It overlies a layer of sand (thickness: $>12.0$ cm)), and the sand sits upon basal gravel. Stratigraphic features of tephras in Sequences E and D at TS (northwest of Mono Lake) and SEB (only Sequence D; southeast of Mono Lake) are listed in Tables \ref{ts_e2c_feature} and \ref{seb_d2c_feature}, respectively.

\subsection{Sequence C (C15-C8)}
Tephras in Sequence C are thicker and coarser than those in Sequences E and D. Some of them display consistent stratigraphic or sedimentologic features at SEB and TS (southeast and northwest of Mono Lake). We list the stratigraphic features of C15-C13 at SEB in Table \ref{seb_d2c_feature}. C12-C8 are described in detail below and in Fig. \ref{c_overall_notext}.

Tephra unit C12 ($7.0$ cm) at SEB (Fig. \ref{c_overall_notext}) is composed of more than five couplets of coarse and thick layers capped by pink, thin laminae of silty to fine-sandy ash. Most coarse layers within C12 are ungraded, massive, gray coarse-sandy ash, but the uppermost one inversely grades from fine-sandy ash to coarse-sandy ash. Their relative thickness is shown in Fig. \ref{c_overall_notext}. The overlying lake silt is $4.5$ cm-thick.

C11 ($18.0$ cm) at SEB is complex-graded (Fig. \ref{c_overall_notext}), and not divided into sub-units during field work. C11 is composed of four, well-sorted, coarse, thick layers, interbedded with five thinner and much finer (silty to fine-sandy ash) layers, the lowermost of which defines the lower contact. Starting from the base, the grain size of the coarse layers is fine lapilli, coarse-sandy ash to fine lapilli, medium lapilli, and coarse-sandy ash to fine lapilli (Fig. \ref{c_overall_notext}). The second and third coarse layers from the base are normally-graded. The overlying lake silt is $6.5$ cm-thick.

C10 at SEB has a total thickness of $21.0$ cm. The basal unit of C10 ($0-<0.5$ cm) is well-sorted coarse-sandy ash.  At $<0.5-\sim4.0$ cm from the base, the unit is well-sorted, and normally grades from fine to medium lapilli to coarse-sandy ash (Fig. \ref{c_overall_notext}). This coarse portion is overlain by a thin pink lamina of silty ash. Above the pink lamina with a sharp contact, there is a $\sim2.0$ cm-thick brown layer that inversely grades from fine-sandy ash to coarse-sandy ash and fine lapilli. A $\sim0.9$ cm-thick layer of planar-laminated, white fine-sandy ash overlies this brown layer. Above this height (half of the total thickness), there are two massive, well-sorted, brown layers of fine-sandy ash to fine lapilli separated by a thin gray lamina of silty to fine-sandy ash with a total thickness of $\sim6.0$ cm. The lower coarse layer is complex graded (fine lapilli to coarse-sandy ash to fine lapilli), while the upper one normally grades from coarse-sandy ash and fine lapilli to coarse-sandy ash. The rest of C10, from $\sim15$ cm above the base to its upper contact, is characterized by numerous pairs of massive white silty ash and gray medium-sandy ash layers with varying thickness. C10 underlies $\sim 14.0$ cm of lake silt.

C10* at SEB is a 0.3 cm-thick layer of black, fine-sandy scoria ash (Fig. \ref{c_overall_notext}). The lake silt overlying it is 18.8 cm-thick. The overlying tephra unit, C9 (9.3 cm), is divided into three sub-units. C9-a (4.5 cm) inversely grades from poorly-sorted fine-sandy ash to fine lapilli; the uppermost few millimeters of C9-a is composed of massive white fine-sandy ash. C9-b (2.9 cm) is composed of two massive, coarse-sandy ash to fine lapilli, gray layers, separated by a thin lamina of planar-laminated silty to fine-sandy ash (0.1 cm). The upper layer is overall finer, and displays inverse grading. C9-c (1.9 cm) is composed of numerous pairs of gray silty ash and fine- to medium-sandy ash, which is similar to the uppermost portion of C10 at SEB. The coarse and fine layers within C9-c are also slightly contorted. The lake silt above C9 is $\sim 62.7$ cm-thick.

C8 (23.1 cm) at SEB is divided into three sub-units (Fig. \ref{c_overall_notext}). C8-a and C8-b (6.7 and 6.4 cm) are two massive, inversely graded, dark gray layers of coarse-sandy ash to medium lapilli. They are separated by a thin lamina of planar-laminated fine- to coarse-sandy ash. C8-c (10.0 cm) is composed of pairs of white, medium-sandy ash and white silty ash layers. Compared with the similar stratigraphy observed in the upper portion C9 and C10, the coarse and fine layers within C8-c are thicker (Fig. \ref{c_overall_notext}). The uppermost part of C8-c displays hummocky cross-lamination and lenses (?), and is slightly contorted.

Tephras in Sequence C at TS (northwest of Mono Lake) are in general thinner than at SEB. Here we describe only characteristics of C11, as it is the thickest Sequence C unit at TS. Features of other tephra units and the interbedded lake sediments within Sequence C at TS are summarized in Table \ref{ts_e2c_feature}.

C11 (14.6 cm) at TS  is divided into six sub-units (Fig. \ref{c_overall_notext}). C11-a (1.8 cm) is massive, pink fine-sandy ash. Overlying sub-units, C11-b and C11-c (0.2 and 0.3 cm), are composed of planar-laminated coarse- and medium-sandy ash, respectively.  C11-d (1.8 cm) is planar-laminated, fine-sandy ash. C11-e (9.7 cm) is the thickest sub-unit of C11. Its grain size ranges from fine-sandy ash to medium lapilli. Within C11-e, there are at least five coarse layers with decreasing thickness interbedded with layers of laminated gray fine-sandy ash. The lowermost coarse layer is rich in obsidian, and normally grades from medium lapilli to coarse-sandy ash. The overlying coarse layer is finer, and inversely grades from coarse sandy-ash to fine lapilli. Contacts between coarse and fine layers in the upper portion of C11-e become more gradual with height (Fig. \ref{c_overall_notext}). C11-f (0.8 cm) is the uppermost sub-unit of C11, and is dark gray, medium- to coarse-sandy ash with slight contortions. The thin tephra layer that directly lies above C11 at TS is reported here as C10, but we stress that further work is necessary to validate this argument. The correlation is made based on matching the number of tephra layers.

At CC (north of Mono Lake), the observed tephra units are C9 and C8. C9 (4.3 cm) lies above lake silt, and is divided into two sub-units (Fig. \ref{c_overall_notext}). C9-a (2.6 cm) has an irregular lower contact, and is composed of six gray layers of sandy-ash, which are separated by thin and massive silty to fine sandy-ash. The lowermost layer inversely grades from silty ash to sandy ash. C9-b (1.7 cm) is planar-laminated, tan to gray, silty to fine-sandy ash.  C8 and C9 are separated by 64.0 cm of lacustrine silt. C8 is a thin white layer (0.2 cm) that normally grades from fine-sandy ash to silty ash. The lacustrine silt and fine sand between C8 and B7 at this site is 17.5 cm-thick.

\subsection{Sequence B (B7*-B5)}
The B7*-B5 tephra layers are sampled at SEB, TS, and WS, and B7 is also found at CC. They are thinner than the Sequence C tephras, and lack internal structures. The stratigraphy of Sequence B tephras not described in the text is summarized in Table \ref{seb_b_feature}. The thin basaltic layer B7* was only found at sites SEB (southeast of Mono Lake; Fig. \ref{a3_photo}a) and WS (northeast of Mono Lake; Fig. \ref{a3_photo}c). At SEB, B7* directly underlies B7, and it is overlain by 1.0 cm-thick lake silt at WS. Features of B7* are given in Table \ref{seb_b_feature} and Fig. \ref{b_overall_notext}. Selected field photos of B7* and B7 in the field are given in Fig. \ref{a3_photo}a-c.

At SEB (southeast of Mono Lake), the thickness of the lacustrine silt that separates Sequences C and B is approximately 160 cm. B7 (5.6 cm) overlies B7*, and can be divided into three sub-units (Figs. \ref{b_overall_notext} and \ref{a3_photo}a). B7-a (3.7 cm) is a massive layer that inversely grades from fine-sandy ash to coarse-sandy ash and fine lapilli. Its color changes from dark brown to white upwards. The grading of the uppermost $\sim1.1 $cm of B7-a is less apparent, but it is slightly normally graded, from coarse-sandy ash and fine lapilli to coarse-sandy ash. B7-b (1.0 cm) is massive, white silty ash with a wavy upper contact. The overlying sub-unit, B7-c (0.8-1.0 cm), is characterized by four couplets of dark brown, coarse sandy scoria ash and white, silty to fine-sandy ash layers. Thickness of each couplet varies from 0.1-0.2 cm. The lower three couplets are wavy and parallel to the lower contact of B7-c, but the uppermost couplet is planar. B7-c has a diffuse upper contact.

At TS (northwest of Mono Lake), B7 (9.6 cm) directly overlies gray lake silt. It is divided into four sub-units (Figs. \ref{b_overall_notext} and \ref{a3_photo}b). B7-a (4.8 cm) is distinctly clast-supported, well-sorted, white, medium-sandy ash to fine lapilli. It shares a sharp but irregular contact with the underlying lake silt, and its uppermost 1.0 cm normally grades to medium-sandy ash. B7-b (2.0 cm) is a complex-graded (fine-sandy ash to medium-sandy ash to fine-sandy ash) massive, white sub-unit with a gradual upper contact. B7-c (1.8 cm) is massive, gray, fine- to medium-sandy ash mixed with dark brown coarse-sandy scoria ash and fine lapilli-sized scoria. B7-d (1.0 cm) is low-angle cross-laminated, white, fine-sandy ash. It has gradual lower and upper contacts with B7-c and the overlying lake silt. There might be a thin and reworked layer that is a few centimeters above B7, but it was not logged during fieldwork. This possibly reworked layer is thus not drawn in the present stratigraphic column in Fig. \ref{b_overall_notext}, but is labeled in Fig. \ref{a3_photo}b with question mark.

The stratigraphy of tephras in Sequence B at WS (northwest of Mono Lake) is similar to that at SEB. B7* and B7 are
separated by 1.0 cm of buff lake silt (Fig. \ref{a3_photo}c). B7 (3.6 cm) is divided into four sub-units (Fig. \ref{b_overall_notext}). B7-a (1.0 cm) is inversely-graded, clast-supported, white coarse-sandy ash. B7-b (1.0 cm) is a thin layer of gray silty ash. The overlying sub-unit, B7-c (0.8 cm), is composed of massive buff silty ash interbedded with at least three planar dark laminae ($\sim 0.1$ cm, each of coarse-sandy scoria). B7-d (0.8 cm) has gradual lower and upper contacts, and is a thin layer of silty to fine-sandy ash. Its yellowish color is probably related to mixing with lake silt. At $\sim <1.0$ cm above the upper contact of B7-d, a reworked or epiclastic layer  ($ < 0.5$ cm) is observed. It is planar-laminated silty to fine-sandy ash, and has gradual lower and upper contacts.

Thickness of the lake silt between B7 and B6 is approximately $25$ cm (?). B6 is 1.3 cm-thick, and composed of five interbedded thin laminae ($\sim0.2-0.3$ cm each) of white fine-sandy ash and gray medium- to coarse-sandy ash, plus a reworked white lamina of coarse-sandy ash on top ($\sim 0.3$ cm; Fig .\ref{b_overall_notext}). This reworked lamina has gradual lower and upper contacts.

B5 at WS has a total thickness of 2.2 cm, and is divided into three sub-units (Fig. \ref{b_overall_notext}). B5-a (0.2 cm) is massive, gray coarse-sandy ash. B5-b ($\sim 0.8$ cm) is massive white fine-sandy ash interbedded with two dark laminae ($\sim0.1-0.2$ cm) of coarse-sandy ash. The laminae are rich in obsidian, and the upper lamina defines the upper contact of B5-b. B5-c ($\sim1.2$ cm) can be further separated into two portions. The lower portion (0-$\sim0.3$ cm from the base) is massive, white fine-sandy ash, while the upper portion (the rest of B5-c) is characterized by low-angle cross-laminated medium-sandy ash. B5-c is overlain by massive lake silt. B7 and B6 were observed at CC (north of Mono Lake), and their stratigraphy is summarized in Table \ref{seb_b_feature}.

\subsection{Sequence A (A4-A1)}
Tephras in Sequence A at proximal sites display complicated layering, which is distinct from the characteristics observed at distal sites. This makes it hard to build up direct correlations between sub-units from site to site. Selected pictures of A4, A3, and A1 in the field are given in Fig. \ref{a3_photo}d-l.

\subsubsection{A4}

A4 (12.6 cm) at HMC (southwest of Mono Lake) is in sharp contact with the underlying and overlying lake sediments (Figs. \ref{a4p} and \ref{a3_photo}d). Its lower portion (7.1 cm) is composed of three distinct sub-units of coarse-sandy ash (A4-a-c with thicknesses of 2.3, 3.2, and 1.7 cm, respectively), which are separated by thin laminae of fine sandy ash. A4-d (3.1 cm) is a complex-graded layer of fine- to medium-sandy ash. The uppermost sub-unit, A4-e (2.4 cm), is white, complex-graded, sandy ash. Wavy and contorted laminae, cross-bedding, and load structures can be observed.

At RC1 (south-southwest of Mono Lake), A4 (29.0 cm) overlies dark gray to brown rounded fluvial gravel deposit ($>4$ m; Fig. \ref{a4p}). A4-a (4.5-5.0 cm) inversely grades from fine-sandy ash to fine lapilli. A thin lamina (0.5 cm) of silty to fine-sandy ash separates it from A4-b (8.3 cm), which is a massive, well-sorted, and normally graded sub-unit of coarse-sandy ash to fine lapilli. The uppermost part of it normally grades to medium-sandy ash with the presence of discrete fine lapilli. Sharing a gradual contact, A4-c ($\sim 5.0$ cm) is mainly composed of massive, gray, fine- to coarse- sandy ash, but its base is rich in discrete fine lapilli.

With a gradual and wavy lower contact, A4-d (2.0 cm) is a massive sub-unit of well-sorted fine lapilli, although mixing with fine- to coarse-sandy ash can be observed at its base. A4-e (8.7 cm), the uppermost sub-unit, is characterized by massive to planar (lower half) and contorted (upper half) laminations of gray silty ash. Discrete white fine lapilli is preserved within it. They are concentrated at $\sim1.5$ and $\sim$ 6.0 cm above the base. For the upper part of A4-e, individual fine lapilli-sized pumice forms a single discontinuous and strongly contorted streak that cuts across the laminations of silty ash (Fig.\ref{a4p}). Despite the presence of contorted lamination, A4-e has a sharp and wavy lower contact, and shares a sharp and planar upper contact with the overlying laminated lake silt. 

At ND2 (south-southwest of Mono Lake), the deposit below A3 is pyroclastic down to 0.7 m below the base of A3. Three sub-units are recognized, and considered to be related to A4 (Fig. \ref{a4p}), but it remains hard to tell if they are primary pyroclastic deposits, or epiclastic, from erosion and weathering processes. Details on the stratigraphy of A4 at ND2 are shown in Fig. \ref{a4p}. It is worth noting that the uppermost portion of A4 is characterized by strongly contorted laminations of orange, ungraded, fine- to medium-sandy ash, and the laminations are locally cut across by the overlying A3 (Fig. \ref{a4p}). No lake silt was preserved between A4 and A3. A3 at ND2 is poorly sorted, lacks grading, and has an irregular, erosional lower contact.

A4 ($\sim 6.8$ cm) at MC (northwest of Mono Lake) is divided into three sub-units (Figs. \ref{a4p} and \ref{a3_photo}e). A4-a ($\sim2.9$ cm) is massive, gray, coarse-sandy ash. At $\sim 1.6$ cm above the base, there is a thin lamina ($\sim<0.2$ cm) of medium-sandy ash with diffusive lower and upper contacts. The rest of A4-a inversely grades from gray medium-sandy ash to coarse-sandy ash. Another thin pink lamina of fine-sandy ash above A4-a defines its contact with A4-b. A4-b ($\sim2.8$ cm) normally grades from medium-sandy ash to fine-sandy ash. The normal grading is interrupted by a $\sim$0.4 cm layer (with diffusive upper and lower contacts) of gray fine-sandy ash that occurs at $\sim1.3$ cm above the base of A4-b. A4-c ($\sim1.1$ cm) is low-angle cross-laminated, pink, silty ash (see Fig. \ref{a4p}).

At DC (south-southwest of Mono Lake), A4 (20.5 cm) is divided into six sub-units (Figs. \ref{a4d} and \ref{a3_photo}f). A4-a (4.5 cm) is a massive sub-unit that inversely grades from silty ash (laminated) to coarse-sandy ash (massive) and fine lapilli (massive). A4-b (2.0 cm) is a laminated, gray silty ash. It is in gradual and irregular contact with the overlying A4-c (4.5 cm). A4-c is a massive,  gray to black sub-unit of medium- to coarse-sandy ash.  A4-d (1.0 cm) is a massive gray fine-sandy ash. A4-e (6.0 cm) is a massive, normally-graded (slight inverse grading at the base) sub-unit of fine lapilli and medium- to coarse-sandy ash. A4-f (1.0 cm) is planar-laminated gray silty ash. A4-g (1.5 cm) is composed of dark gray medium-sandy ash (0-0.6 cm at the base) overlain by laminated gray silty ash. The overlying laminated lake silt is 5.0 cm-thick.

A4 overlies a yellowish deposit at BC (north of Mono Lake), but its source remains unknown (lake silt or epiclast). A4 (10.25 cm) is divided into five sub-units (Fig. \ref{a4d}). A4-a (2.5 cm) is composed of three portions (starting from the base: 0.5 cm, fine-sandy ash; 0.5 cm, coarse-sandy ash; 1.5 cm, fine-sandy ash). A4-b (2.2 cm) is poorly sorted, rich in lithics, and composed of fine-sandy ash to medium lapilli. A4-c (1.3 cm) is a sub-unit of fine- to medium-sandy ash with its coarse material concentrated at the lower $\sim0.25$ cm. A4-d (2.75 cm) is poorly sorted, and normally grades from medium lapilli to coarse-sandy ash. A4-e (1.5 cm) is a white layer of massive fine-sandy ash. At 0.5-1.0 cm from the base, three distinct gray laminae of medium- to coarse-sandy ash can be observed. A4 is capped by a hardened calcite shelf.

A4 (7.0 cm) at WS (northeast of Mono Lake) can be divided into four sub-units (Fig. \ref{a4d}). A4-a (2.0 cm) is a massive, obsidian-rich sub-unit, and its grain size ranges from coarse-sandy ash to fine lapilli. Its uppermost portion (starting from $\sim$0.2 cm below its upper contact) grades to gray fine-sandy ash. A4-a has a gradual upper contact with A4-b (1.8 cm), which is a massive yellowish sub-unit of silty ash. The stratigraphy of A4-c (2.1 cm) and A4-d (1.1 cm) are highly similar to A4-a and A4-b respectively. The only difference is that there is a gray thin lamina ($\sim0.6$ cm above the lower contact of A4-d) of medium-sandy ash with a thickness of $\sim0.25$ cm preserved within A4-d (Fig.\ref{a4d}).

\subsubsection{A3}
A3 has been observed at all sample sites. Stratigraphic columns of A3 at HMC, RC1, and ND1 are shown in Fig. \ref{a3p}. At HMC (southwest of Mono Lake), A3 ($\sim 9.6$ cm) is divided into four sub-units. A3-a ($\sim1.8$ cm) is a massive tan fine- to medium-sandy ash with rich ostracod shells. A3-b ($\sim 5.5$ cm) is white inversely-graded medium-sandy ash to fine lapilli with gradual lower and upper contacts. A3-c ($\sim 2.3$ cm) is massive, normally-graded, gray, fine- to medium-sandy ash to silty ash with low-angle cross-lamination observed. It underlies $2.2$ cm-thick laminated gray lake silt.

A3 (50.2 cm) at RC1 (south-southwest of Mono Lake) is divided into eight sub-units that display complex structures (Fig. \ref{a3p}). The base sub-unit A3-a (1.0 cm) is poorly-sorted, and composed of medium- to coarse-sandy ash to fine lapilli. A3-b (0.8 cm) is massive, gray, fine-sandy ash. Ostracod shells and biotite are abundant in the sample of A3-a and A3-b.

A3-c is in slight irregular contact with A3-b, and the contact is a thin pink layer of silty ash. A3-c (3.6 cm) is a loose, normally-graded, yellowish sub-unit with grain size ranging from medium-sandy ash to medium lapilli. The sorting is poor at the base, but it improves upwards with the depletion in fine-sandy ash (note that this does not conflict its overall normal grading). The upper portion of this sub-unit is characterized by imbrication or parallel alignment of coarse-sandy ash and fine lapilli. The contact between A3-c and A3-d is characterized by an abrupt change in color (yellow-gray). A3-d (4.0 cm) normally grades from coarse- to fine-sandy ash with wavy lamination. At its uppermost portion, this sub-unit turns yellowish.

A3-e (6.1 cm) is white in color, and is characterized by convolute laminations of fine- to medium-sandy ash. A small cone-shaped deposit (at decimeter scale), perhaps formed due to the ejection of underlying deposit as a result of liquefaction (known as sand blow), is preserved in between A3-d and A3-e (Fig. \ref{a3p}). There are clear boundaries that further divide A3-e into three portions based on their different colors (starting from the base: white, gray, and white), which are also related to their grain size (fine-sandy ash, medium-sandy ash, fine-sandy ash). There are discrete coarse-sandy ash to fine lapilli-sized pumice grains that are preserved exactly along the contact of the second (gray) and third (white) portions. Contacts of this sub-unit and the ones between the three portions are in general parallel to the contour of the cone-shaped deposit below them. A centimeter-scale ramp-thrust fault that cuts the cone-shaped deposit as well as the lower portion of A3-e is also observed.

A3-f (29.0 cm) is the thickest sub-unit of A3 at RC1. Its grain size ranges from fine-sandy ash to medium lapilli. A3-f inversely grades from fine-sandy ash to medium lapilli to $\sim 9.6$ cm above the lower contact. This portion of the deposit is also characterized by parallel alignment of medium lapilli. At $\sim 9.6$ cm above the lower contact, A3-f turns to massive coarse-sandy ash, and then gradually grades to planar-laminated medium- and fine-sandy ash. Starting from $> \sim 3.0$ cm below its upper wavy contact, the uppermost portion of A3-f is composed of hummocky cross-laminated gray to white fine- to medium-sandy ash. A burrow is observed here which intrudes A3-f. A3-g (5.0 cm) is a massive sub-unit of white silty- to fine-sandy ash. It has sharp contacts with A3-f (wavy) and A3-h (0.7 cm; planar). A3-h is the uppermost sub-unit of A3 at RC1. It is massive, gray, fine- to medium-sandy ash, and underlies $19.3$ cm-thick lake silt. At RC2, four sub-units of A3 are recognized, which are highly consistent with sub-units A3-f to A3-h at RC1.

At ND1 (south-southwest of Mono Lake), A3 (85 cm) is a single, normally graded unit with grain size ranging from medium lapilli at the base to medium- to fine-sandy ash on top (Fig. \ref{a3p}). Rhyolitic lithics are abundant in the lower 30.0 cm, which is also poorly-sorted. Its upper contact is gradual.

A3 (40.1 cm) at DC (south-southeast of Mono Lake) has four sub-units (Fig. \ref{a3d}). A3-a (4.5 cm) is a massive, dark gray, fine ungraded lapilli. A3-b (2.5 cm) is a massive sub-unit that inversely grades from silty ash to sandy ash with a gradual upper contact.  A3-c (13.0 cm) is a massive, and complex-graded sub-unit (fine-sandy ash to coarse-sandy ash and fine lapilli). The lower half is normally graded, and composed of fine lapilli and coarse-sandy ash (slight inverse grading at the gradual lower contact). In the central portion of A3-c, the deposit is mainly sandy ash mixed with fine lapilli. The upper one third of A3-c consists of massive, non-graded, gray fine lapilli with a minor fraction of coarse-sandy ash concentrated below the upper contact. A3-d (21.0 cm) is the thickest sub-unit of A3 at DC. It is characterized by an alternation of planar or wavy coarse and fine layers (see Fig. \ref{a3d} for more details). The coarse layers are composed of coarse-sandy ash and fine lapilli (minor), and the fine layers are massive, gray silty ash.

At MC (northwest of Mono Lake), A3 ($\sim 3.10$ cm) can be divided into six sub-units (Figs. \ref{a3d} and \ref{a3_photo}g). A3-a (0.40 cm) is composed of pink-silty to fine-sandy ash. The overlying sub-unit, A3-b (1.00 cm), is planar-laminated, gray, medium-sandy ash. A3-c ($\sim 0.8$ cm) is a massive pink sub-unit of silty to fine-sandy ash. A3-d and A3-e are similar in thickness ($\sim 0.35-0.45$ cm) and stratigraphy. Their color turns from black to pinkish upwards perhaps due to the decrease in biotite and some obsidian grains. The color change is more rapid for A3-e. A3-d and A3-e are also characterized by a slight normal grading from silty to fine-sandy ash to silty ash. A3-f ($\sim$0.15 cm) is composed of massive, gray, fine-sandy ash.

A3 at BC (north of Mono Lake) is divided into six sub-units with a total thickness of $\sim 4.6$ cm (Figs. \ref{a3d} and \ref{a3_photo}h). Most of them have no obvious grading. It has a slight irregular contact with the underlying lake silt. At the contact, a discrete layer of coarse-sandy to fine lapilli-sized lithics and obsidian can be observed. A3-a (1.6 cm) is massive silty ash. A3-b (0.3 cm) is in slight diffusive contacts with sub-units below and above. It is massive, dark gray, fine- to medium-sandy ash. A3-c (0.5 cm) is planar-laminated, pinkish, silty ash. A3-d (0.6 cm) is characterized by the presence of dark  fine- to medium-sandy ash at the base. A3-d normally grades to silty ash upwards with color changing to pink accordingly. A3-e ($\sim$0.7 cm) shares a similar stratigraphy with A3-d. The only difference is that the change in grading, color, and grain size is more rapid, and occurs only at the base. Complex grading can be observed at A3-f ($\sim$0.9 cm), the uppermost sub-unit of A3 at BC. It is mainly composed of pink silty ash. Three dark laminae of fine-sandy ash are interbedded within it at the lower two thirds of A3-f. The first lamina defines its lower contact. The three laminae have a comparable thickness ($\sim 0.1$ cm), and are in diffusive contacts with the underlying and overlying deposits.

A3 at WS (northeast of Mono Lake) is mainly characterized by complex grading with alternations of massive silty to fine-sandy ash and dark laminae of medium- to coarse-sandy ash. We separate them into three sub-units for simplicity (Figs. \ref{a3_photo}i). A detailed stratigraphic column of A3 at WS can be found in Fig.\ref{a3d}. A3 at WS has a total thickness of 6.3 cm. With an irregular lower contact, the base sub-unit, A3-a (1.2 cm), is rich in obsidian ($\sim20-30\%$ from field measurement), and inversely grades from medium-sandy ash to fine lapilli. A3-b ($\sim3.6$ cm) is complex graded, and composed of six dark laminae interbedded with yellowish silty to fine-sandy ash. The grain size of these laminae is characterized by abundant aggregates with grain size ranging from medium- to coarse-sandy ash. Since the laminae are loosely packed and therefore little modified by secondary processes, the aggregates observed are assumed to be primary depositional structures. A3-c ($\sim 1.5$ cm) is massive, yellowish, silty ash.

\subsubsection{A2}

The stratigraphy of A2 at outcrops close to its source, the Black Point Volcano, has been widely studied in previous works \citep{custer1973stratigraphy,white2000subaqueous,murtagh2013pyroclast}. At outcrops except for MC, A2 is not thicker than 5.0 cm, but can be recognized due to its distinct dark brown to black color. A2 at these sites is composed of two to four primary sub-units separated by lake silt, plus one or two underlying or overlying, thin, and possibly reworked layers. The reworked layers above the primary sub-units of A2 are identified based on their lighter color. Most primary deposits of A2 are poorly sorted and non-graded. Notably, A2 and (or) the underlying and overlying deposits are contorted at HMC, RC2, and ND2 (See Fig. \ref{a2o}). Their thickness, grain size, and other features (e.g., which sub-layer is reworked) are also summarized in Table \ref{a2_summary}.

\subsubsection{A1}
A1 was found at all sample sites, except for BC and RC2 due to erosion. At HMC (southwest of Mono Lake), A1 (16.9 cm) is divided into three sub-units (Fig. \ref{a1p}). A1-a ($\sim 5.2$ cm) is complex-graded, planar-laminated white to light gray medium- to coarse-sandy ash. It is characterized by six pairs of coarse and fine layers with comparable thickness. Lake silt (2.2 cm) divides A1-a and A1-b. A1-b (7.0 cm) is complex-graded, laminated, medium- to coarse-sandy ash and fine lapilli. It is white to gray in color, and rich in obsidian. Two white laminae of fine-sandy ash can be observed at $\sim0.9$ and $\sim4.2$ cm above the base. A1-c (2.5 cm) is laminated, white to gray, fine- to medium-sandy ash.

At RC1 (south-southwest of Mono Lake), a light brown layer (22.0 cm) of ash-sized pyroclastic material (grain size $<2.0$ mm) underlies A1. A1 ($\sim36.2$ cm) is separated into nine sub-units (Fig. \ref{a1p}). A1-a-e are composed of couplets of brown to gray layer of coarse-sandy ash overlain by thinner gray layer of silty ash (see Fig. \ref{a1p} for more details). Lake silt ($\sim1.9$ cm) overlies A1-e. Above the lake silt, A1-f ($\sim$ 3.9 cm) is composed of planar-laminated gray fine-sandy ash. The lower half of it is lighter in color. A1-g ($\sim$15.6 cm) is a thick, gray sub-unit with complex grading. The lower one third of A1-g is massive, gray coarse-sandy ash, which is overlain by a thin layer of gray fine- to medium-sandy ash. Above this thin layer, A1-g becomes coarser, and is characterized by massive gray fine lapilli with a slight normal grading. A thin lamina of fine- to coarse-sandy ash can be observed above it (Fig. \ref{a1p}). The uppermost $\sim 3$ cm of A1-g normally grades to medium- to coarse-sandy ash.

A1-h ($\sim1.4$ cm) is composed of gray fine-to medium-sandy ash with complex grading (starting from the base: medium-sandy ash to fine-sandy ash to medium-sandy ash). A1-i ($\sim1.0$ cm) is the uppermost sub-unit of A1 at RC1. It is planar-laminated (locally low-angle cross-laminated), buff, silty ash.

A1 overlies dark green lake silt at ND2 (south-southwest of Mono Lake; Figs. \ref{a1p} and \ref{a3_photo}j). With an irregular contact, the base sub-unit A1-a ($\sim9.0$ cm) is thick, massive, gray silty ash. At its upper portion, the deposit displays several discrete layers ($\sim$1.0 cm) of fine sandy ash with lighter color. A1-a here seems to be correlated with the pyroclastic layer that underlies A1 at RC1. A1-b ($\sim11.7$ cm) is composed of couplets of coarse and thick, and fine and thin layers. At least ten couplets can be recognized (Fig. \ref{a3_photo}j). Coarse layers within the couplets are  massive, non-graded, coarse-sandy ash to medium lapilli, while fine layers are all composed of gray silty to fine-sandy ash, and are wavy (wavelength at centimeter scale) or planar (Fig. \ref{a3_photo}j). A1-c ($\sim1.4$ cm) is composed of cross-bedded gray silty ash mixed with coarse-sandy ash overlain by wavy-laminated gray silty ash. It shares a wavy, sharp contact with the overlying massive lake silt (2.4 cm). A1-d (16.0 cm) with complex grading overlies the lake silt. The lowermost $<0.5$ cm of A1-d is a thin, wavy-laminated layer of pink silty ash. The main body of A1-d is composed of massive, well-sorted, obsidian-rich coarse-sandy ash intercepted by two or three gray thin laminae ($\sim 0.1$ cm) of planar silty to fine-sandy ash (their relative positions and thicknesses are plotted in Fig. \ref{a1p}). The overall grain size of A1-d increases upwards. The uppermost 2.0 cm of A1-d is complex-graded (gray coarse-sandy ash to white fine-sandy ash to gray coarse-sandy ash), which resembles A1-h at RC1. Its upper contact is sharp, but not entirely planar. The uppermost sub-unit of A1 at ND2, A1-e ($1.2$ cm), is low-angle cross-laminated pink silty ash, and is overlain by lake silt.

A1 at MC (northwest of Mono Lake) is characterized by complex grading, and overlies yellowish brown sand with a sharp wavy contact. It is divided into three sub-units, and has a total thickness of 11.1 cm (Figs. \ref{a1p} and \ref{a3_photo}k). A1-a ($4.5$ cm) is characterized by complex-graded, silty to coarse-sandy ash. There are five thin, dark layers of gray, medium- to coarse-sandy ash interbedded with massive, pink, silty ash (Fig. \ref{a1p}), and the lowermost $\sim 1.3$ cm of this sub-unit is also wavy, and parallel to the sharp lower contact. Except for the lowermost gray, coarse layer, which is normally graded with a gradual change in color, the rest thin layers are in sharp contact with the planar-bedded, silty ash below and above. A1-b ($< \sim 5.4$ cm) is a gray, complex-graded sub-unit, and its lower portion is later recognized as lake silt. A thin pink lamina of fine-sandy ash can be observed at $\sim 0.7$ cm below its upper contact. A1-c (1.1 cm) is a planar-bedded pink sub-unit with massive silty to fine-sandy ash. Two thin dark laminae of coarser material can be observed within it (Fig. \ref{a1p}). The deposit above A1-c is lake sediment of silt and fine sand in grain size. During a revisit to MC in 2016, another outcrop (MC-2016; Fig. \ref{a3_photo}l) of A1 near MC was found, whose stratigraphy is highly consistent with A1 at MC.

At MC-2016, A1 is in wavy contact with the underlying lacustrine deposit, which displays low-angle cross-bedding. Below the lake silt (planar-laminated) within A1, all tephra sub-units (corresponding to A1-a at MC) are characterized by wavy-lamination. Tephra sub-units above the lake silt are massive or planar-laminated.

At WS (northeast of Mono Lake), A1 is divided into two massive sub-units, and sits upon gray to green lacustrine silt. A1-a (1.6 cm) is normally graded, gray to pink, silty to fine-sandy ash. A1-b (2.8 cm) is a complex-graded (silty ash to medium-sandy ash to silty ash) ash layer. We suspect that the base of A1-b is composed of lake silt (Fig. \ref{a1p}). From 0-1.9 cm above the upper contact of A1,  the overlying lake deposit is silt to sand in grain size. Further above, the lake deposit turns to laminated silt.

\section{Characteristics of the tephra samples in the WCF}
Commonly observed components ($\le 1 \phi$) of the rhyolitic tephras in the WCF are pumice, obsidian, lithics, and aggregates. Almost all rhyolitic pumice samples are angular to sub-angular tube pumice, and are denser than water. Low-density pumices (lighter than water) are preserved within B7-d at WS and CC, A3-a and A3-d at DC. Within B7-d, sparse, well-rounded, flat, lapilli-sized pumices ($-2 \phi$) are mixed with silty to fine-sandy ash. For A3, we have found only two low-density pumice grains in each sub-unit. They are the coarsest in the samples. For other sub-units, low-density pumices are extremely rare, and do not display distinct features (grain size, roundness, and color) compared with other heavier pumices. Aggregates are commonly observed at distal sites, but in many cases it remains hard to tell if they are primary, or the result of secondary induration.

We have found abundant ostracod shells of varying size in samples of D17 at TS, A3-a at HMC and A3-a-b at RC1. The fact that these shells are well-mixed within the corresponding, lowermost sub-unit, yet they are discrete and sparse in the directly underlying lacustrine deposit suggests that they were deposited together with the tephra after being excavated from more ostracod-rich layers within the lacustrine sequence \citep{L68}. Similar phenomenon has been recognized for the initial stage of the Oruanui eruption of Taupo Volcano \citep{van2013high}.

\begin{figure}[H]
	{\includegraphics[width=\textwidth]{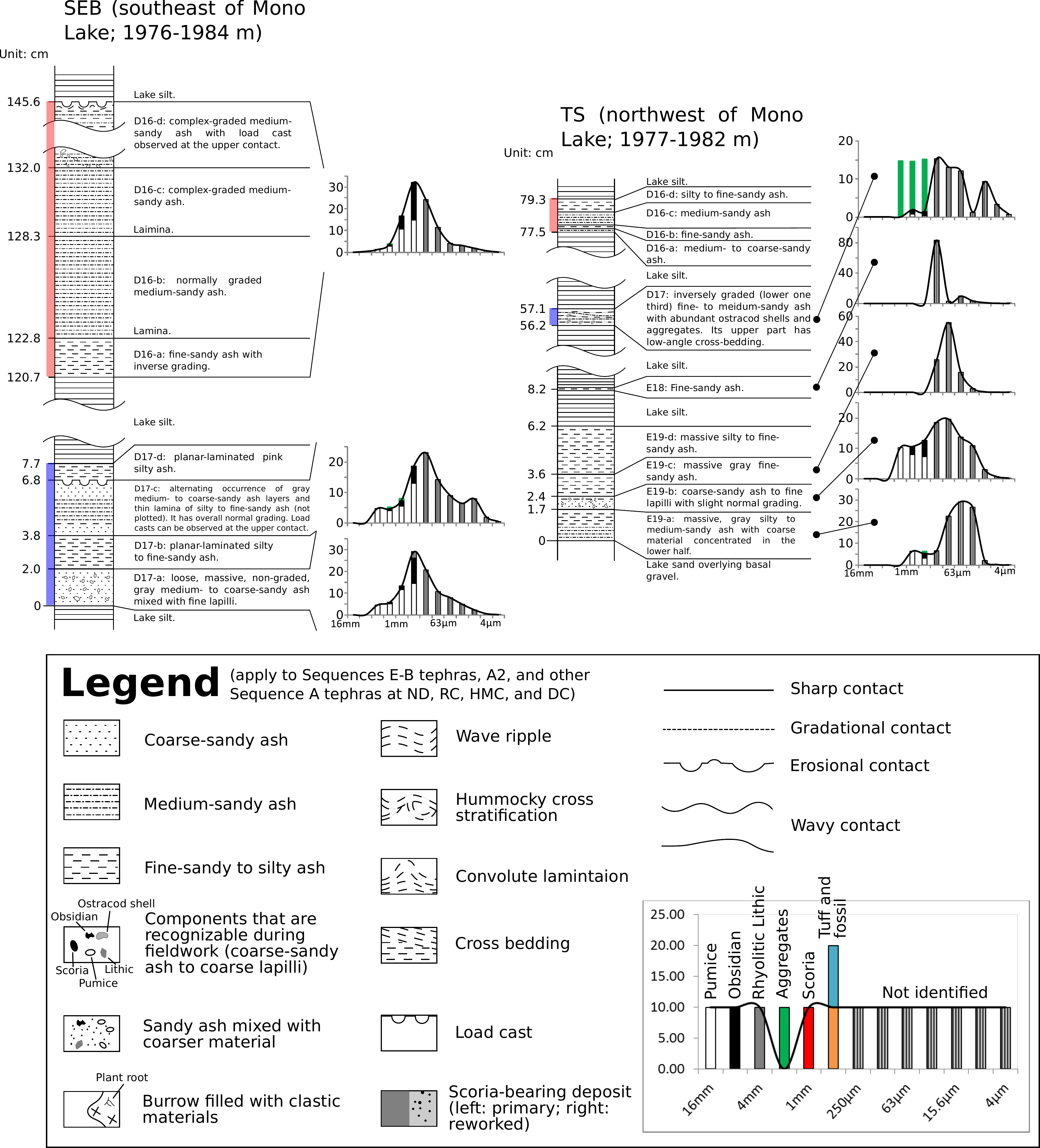}}
	\caption{Stratigraphic columns of tephras in Sequences E and D at TS and SEB (only Sequence D at SEB), and grain size distributions of sampled tephra. Explanation of symbols for stratigraphic columns of Sequences E-B tephras, A2, and other Sequence A tephras at ND, RC, HMC. and DC in this work is shown in this figure. Color bars on the left of the stratigraphic columns denote the correlation between tephra unit (not sub-unit correlation) at different sample sites. The elevation range of the observed tephras at each sample site (as shown in Fig. \ref{ggb}b) is given in bracket. }
	\label{d_overall}
\end{figure}

\begin{figure}[H]
	{\includegraphics[width=\textwidth]{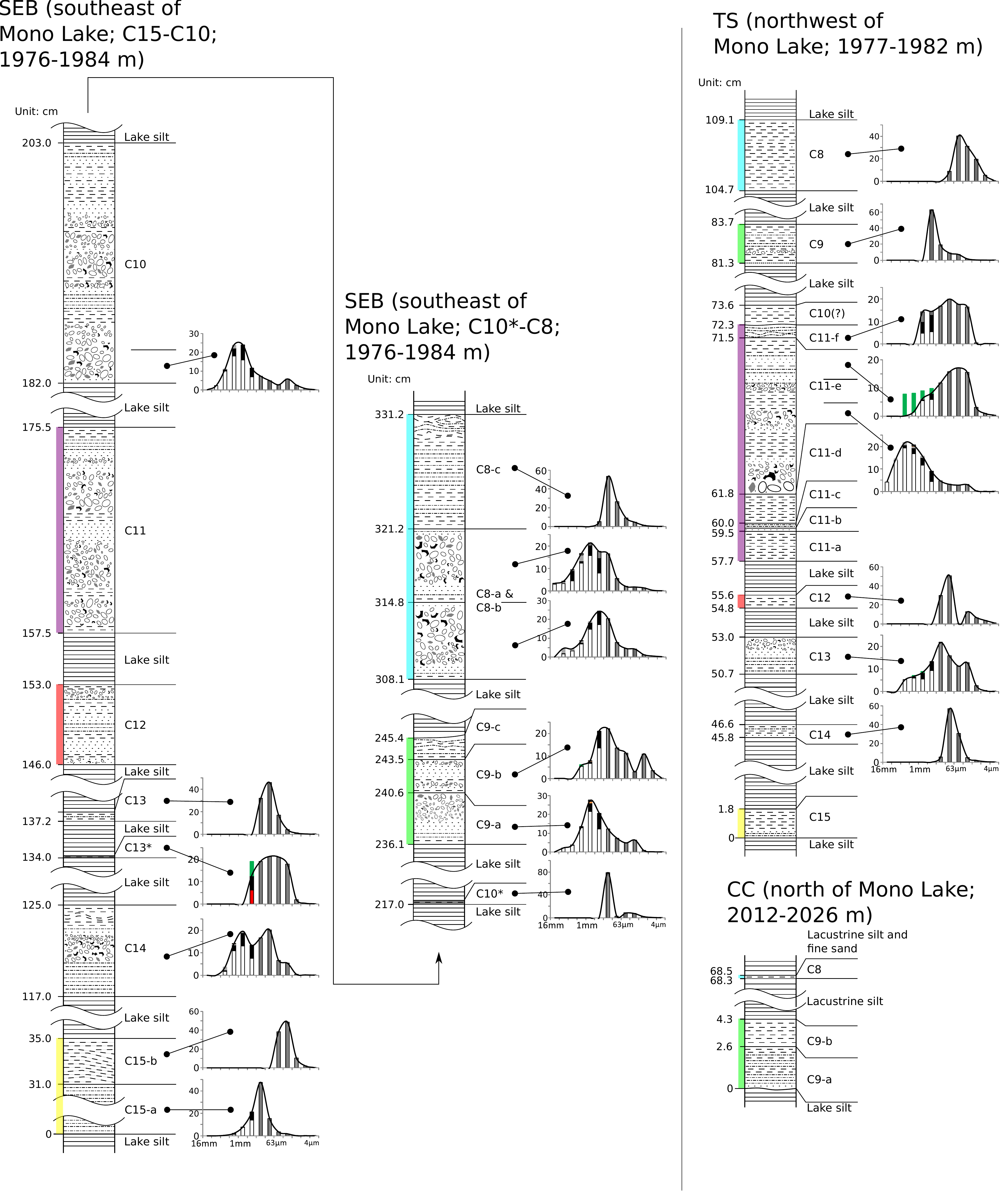}}
	\caption{Stratigraphic columns of tephras in Sequence C at SEB, TS, and CC, and grain size distributions of sampled tephra. See Fig. \ref{d_overall} for explanation of symbols. Color bars on the left of the stratigraphic columns denote the correlation between tephra unit (not sub-unit correlation) at different sample sites. The elevation range of the observed tephras at each sample site (as shown in Fig. \ref{ggb}b) is given in bracket. }
	\label{c_overall_notext}
\end{figure}

\begin{figure}[H]
	{\includegraphics[width=\textwidth]{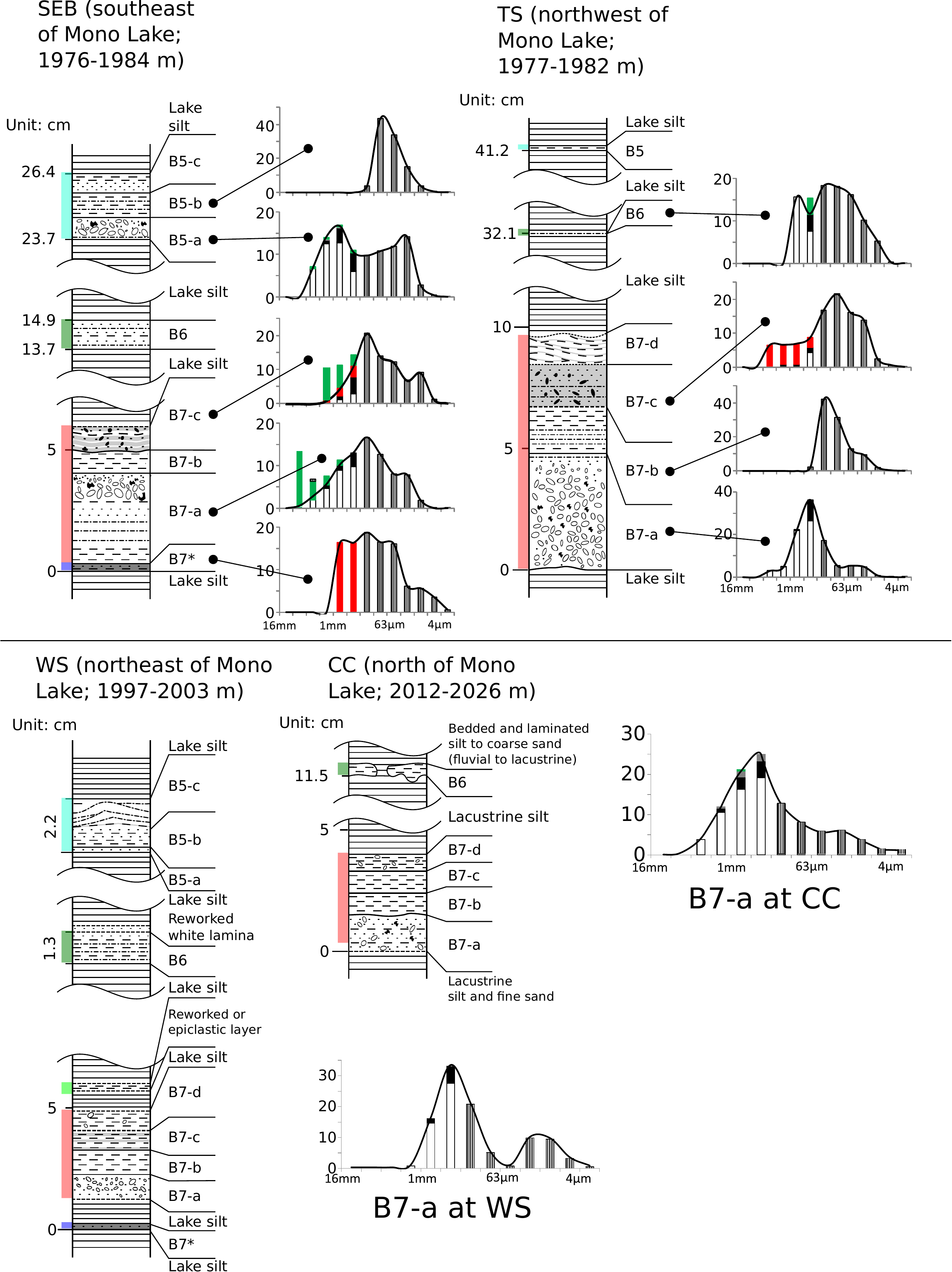}}
	\caption{Stratigraphic columns of tephras in Sequence B at SEB, WS, TS, and CC, and grain size distributions of sampled tephra. See Fig. \ref{d_overall} for explanation of symbols. Color bars on the left of the stratigraphic columns denote the correlation between tephra unit (not sub-unit correlation) at different sample sites. The elevation range of the observed tephras at each sample site (as shown in Fig. \ref{ggb}b) is given in bracket. }
	\label{b_overall_notext}
\end{figure}

\begin{figure}[H]
	{\includegraphics[width=\textwidth]{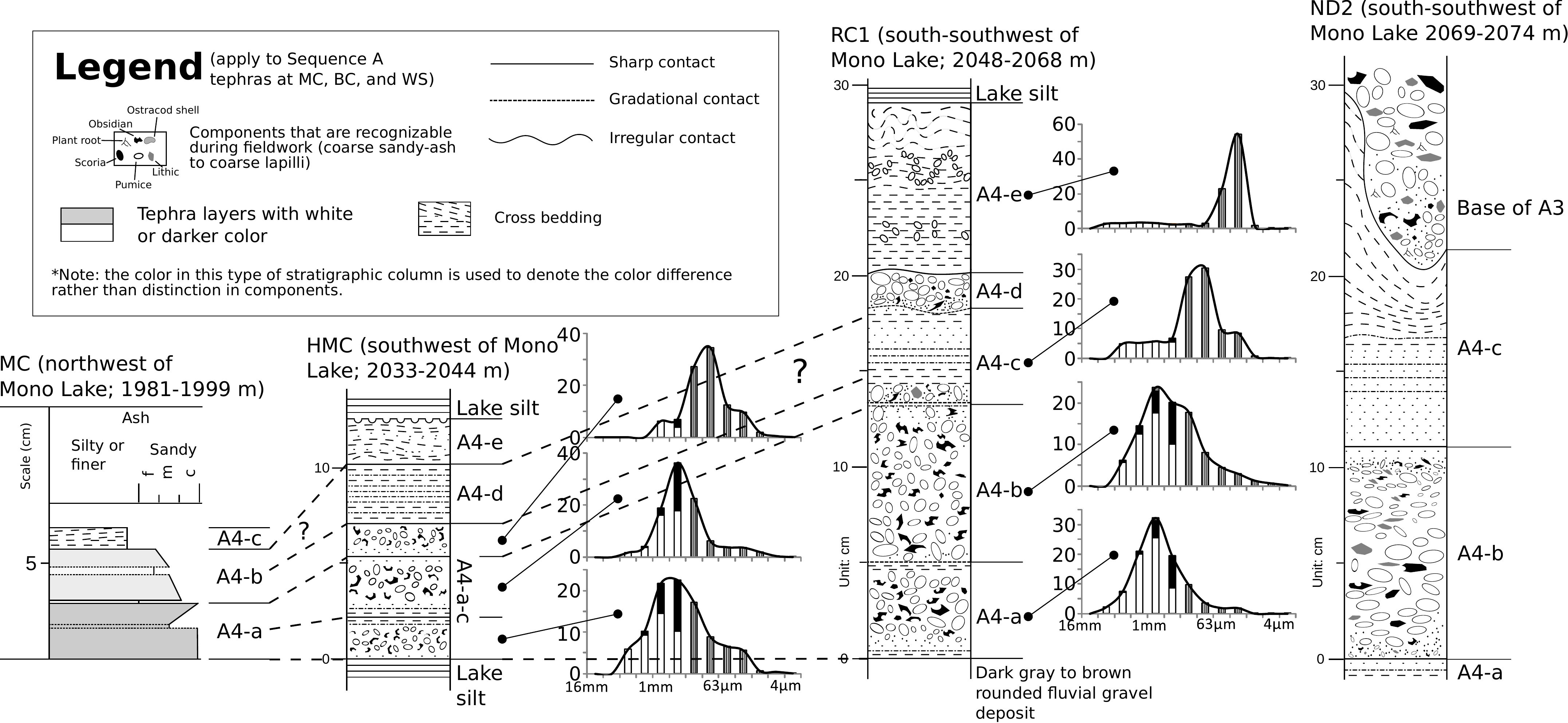}}
	\caption{Stratigraphic columns of A4 at MC, HMC, RC1, ND2, and grain size distributions of sampled tephra. These outcrops are plotted together as they are all in the west of the Mono Craters. Proposed sub-unit correlation is marked with dashed lines. Note that the stratigraphic column of A4 at MC is drawn in a different style. That is because the deposit there is thinner and finer, and it is important to highlight the differences in grain size and color between sub-units. See Fig. \ref{d_overall} for explanation of symbols for stratigraphic columns of A4 at HMC, RC1, and ND2. Explanation of symbols for stratigraphic columns of A4, A3, and A1 tephras at MC, BC, and WS is shown here. The elevation range of the observed tephras at each sample site (as shown in Fig. \ref{ggb}b) is given in bracket. }
	\label{a4p}
\end{figure}

\begin{figure}[H]
	{\includegraphics[width=\textwidth]{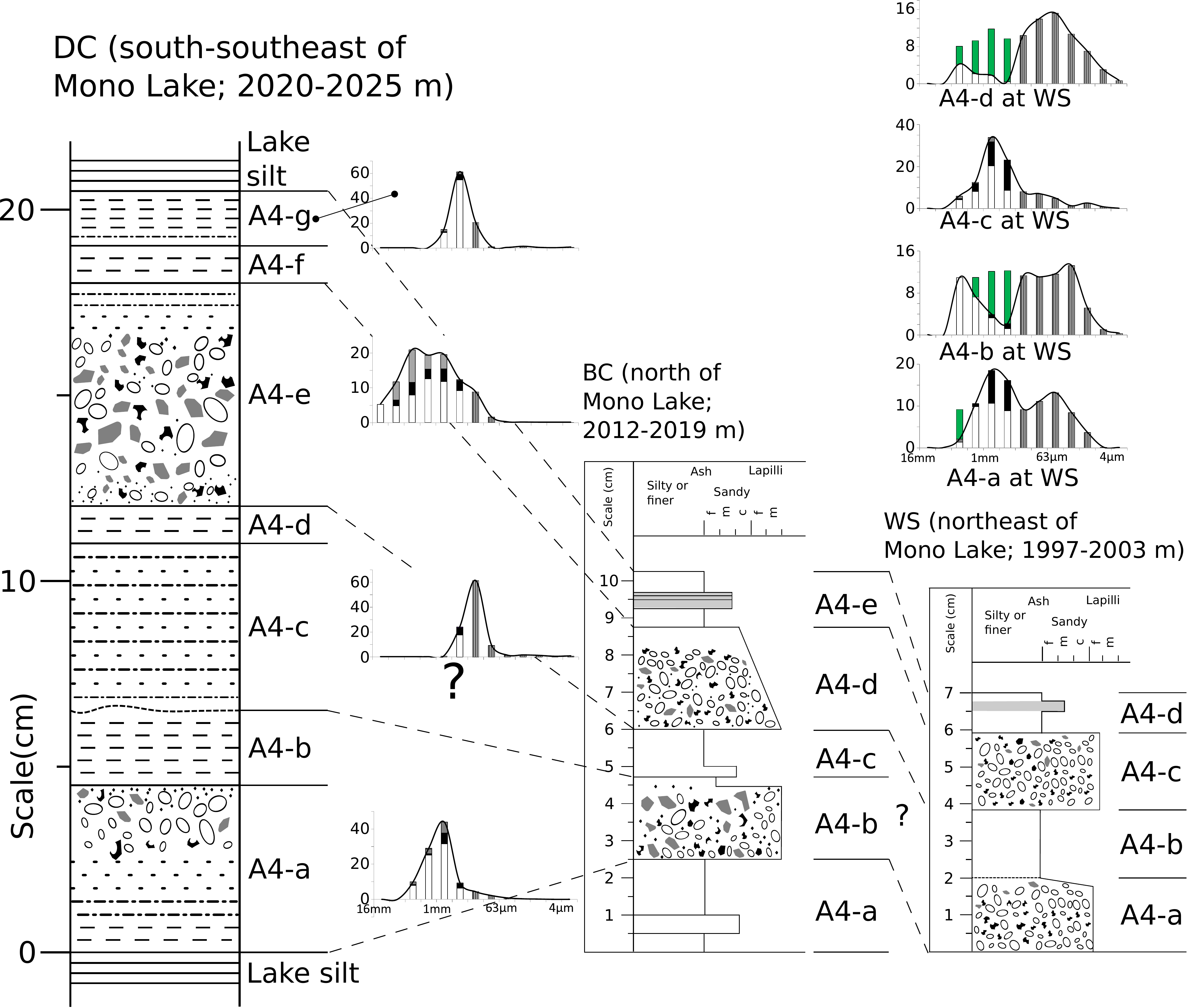}}
	\caption{Stratigraphic columns of A4 at DC, BC, WS, and grain size distributions of sampled tephra. These outcrops are plotted together as they are all in the east of the Mono Craters. Proposed sub-unit correlation is marked with dashed lines. Note that two styles (DC; BC and WS) of stratigraphic column drawing are used in this figure. That is because the deposits at distal sites are much thinner and finer, and it is important to highlight the differences in grain size and color between sub-units. See Figs. \ref{d_overall} and \ref{a4p} for explanation of symbols for A4 at DC and BC and WS, respectively. The elevation range of the observed tephras at each sample site (as shown in Fig. \ref{ggb}b) is given in bracket. }
	\label{a4d}
\end{figure}

Grain size distributions of the sampled WCF tephras are shown together with their stratigraphic columns in Figs. \ref{d_overall}-\ref{a1p}. During field work, samples were collected for an entire tephra unit, a single sub-unit, or a portion of a sub-unit as marked in the figures. Therefore it is possible that characteristics in the grain size distribution are partly caused by channel sampling. They are still presented here as they provide quantitative constraints on the grain size and components of the sampled tephras. With the sub-unit correlation established, the present data could help future workers to infer the eruption parameters of the WCF tephras. In the text, our discussion is based on samples that represent the product from a single eruption pulse, either a single massive unit, or a designated sub-unit.

Grain size distributions shown in Fig. \ref{d_overall}-\ref{a1p} are in general consistent with field observations. It is worth noting that B7-a shows bimodal (WS) and fine-tail skewed (SEB, TS, and CC) grain size distributions. The consistent stratigraphy for B7-a as well as other B7 sub-units at the four sample sites indicates their correlation. B7-a lacks grading (except at SEB), and is distinctively clast-supported. Together with the fine-tail skewing, these observations suggest excess deposition of fines during a sustained, non-waning event. The fines are most likely to be produced from the disintegration of aggregates or the abrasion of pumice grains while floating above water given the present observations. In addition to B7, we notice that A4-e at RC1 and A1-a at ND2 are both characterized by highly coarse skewed grain size distribution (Figs. \ref{a4p} and \ref{a1p}), which is unique among all WCF tephra sub-units.

Some sub-units are characterized by a higher fraction of obsidian or felsite lithics. Under the microscope, features of these felsite lithics (e.g., flow banding) suggest a rhyolitic composition.  Higher fractions of obsidian or rhyolitic lithics are taken to be indicative of water-magma interaction \citep{wohletz1983mechanisms} or excavation of existing domes, respectively, during the eruption.

Most fine tephra samples are composed of pumice with low vesicularity (e.g., Fig. \ref{sem}e), and glass shards (e.g., Fig. \ref{sem}g and h). Samples from a few sub-units display distinct features. E18 is highly vesicular, with roughly equant bubbles (Fig. \ref{sem}a), which is perhaps related to the finding that it was from a different source region, Mammoth Mountain \citep{M14}. Samples from the upper portion of C12 at SEB have smooth surfaces with low angularity (Fig. \ref{sem}b), which suggests that they may be the product of ductile fragmentation, as observed in experiments \citep{buttner2002thermohydraulic,austin2008phreatomagmatic}; Quenching cracks can be observed in samples of C10* (Fig. \ref{sem}c). Such cracks are often produced from water-magma interaction with excessive water \citep{buttner1999identifying, buttner2002thermohydraulic, austin2008phreatomagmatic}. These observations can be used to identify the eruption style in some cases, but we stress that they need to be dealt with caution, since it is necessary to take different lines of evidence into account to identify the eruption style as pointed out by \cite{white2016magmatic}.

\section{Discussion}

In this section, we propose sub-unit correlations for tephras that display similar characteristics at different outcrops. Noticeable stratigraphic features of the WCF tephras are listed. Since sub-units of B7, and tephras in Sequence A, are well-correlated at more than three sample sites, we make interpretations of their eruptive history. Then implications from our work, including eruption style and temporal variation of volcanic activities from the Mono Craters and water level history of Lake Russell during the late Pleistocene, are discussed and summarized.

\subsection{Sub-unit correlation}
The sub-unit correlations proposed in this work are based on the consistency in stratigraphic position and features, such as sorting, grading, and color, of different sub-units. Establishing sub-unit correlation based on physical characteristics is more efficient and practical compared to geochemical methods given the large number of sub-units for the WCF tephras and the potential chemical homogeneity between tephra sub-units.  The sub-unit correlations could help further constrain the correlation between the WCF tephras and their source vent, as well as with other contemporary ashes deposited outside the Long Valley-Mono Craters region. Tephras in Sequence E are observed only at TS, and tephras in Sequence D are thin at TS. Therefore no sub-unit correlation can be made for them. The same argument also applies to tephras C15, C14, C12, C10, and C8.

C13 is a thin unit at both SEB (0.8 cm; southeast of Mono Lake) and TS (2.3 cm; northwest of Mono Lake). If deposits at the two sites (Fig. \ref{c_overall_notext}) are products from different pulses of the eruption, distinct sedimentologic characteristics at the two sites are expected. However, the lower two-thirds of C13 at SEB and the whole of C13 at TS are both inversely graded.  Given the inverse grading, and the fact that the stratigraphy of C13 at each site represents tephra deposited from at most two eruption pulses, we propose the correlation between the lower two-thirds of C13 at SEB and the whole of C13 at TS.

C11 is the thickest tephra unit in Sequence C, and its main feature, observed at both SEB and TS, are the four to five couplets of coarse lapilli and fine ash layers (C11 at SEB and C11-e at TS; Fig. \ref{c_overall_notext}). The coarse lapilli-bearing lower layers are rich in obsidian and subangular pumice at both sites, suggesting high columns during the eruptions. As the thickest and coarsest sub-units within Sequence C observed at the two sites, C11 at SEB and C11-e at TS are hence correlated. C11 at SEB lacks other finer and thinner sub-units of C11 exposed at TS.

C9-a and C9-b at SEB have a total thickness of 7.4 cm, and C9 is 2.4 cm-thick at TS. C9-a and C9-b at SEB, taken together, can be regarded as comprising a triplet, with the lowermost and uppermost parts having inverse grading (Fig. \ref{c_overall_notext}). This description is consistent with the stratigraphy of C9 at TS, except that the inverse grading for the extremely thin ($\sim 0.8$ cm), uppermost portion of C9 at TS is not observed. Following the argument made for C13, the probability that these thin deposits were derived from distinct eruption pulses while sharing consistent stratigraphy at the two sites is small.  Correlation is thus proposed between C9-a and C9-b at SEB with the whole of C9 at TS.

Tephras in Sequence B have been observed at three to four sites. Since SEB (southeast of Mono Lake) and WS (northeast of Mono Lake) are nearly in the same direction from the Mono Craters, Sequence B tephras at the two sites are consistent in stratigraphy (Fig. \ref{a3_photo}a and c). Three features of B7 establish a well-constrained sub-unit correlation at the four sample sites. These are the massive, well-sorted coarser material (coarse-sandy ash to fine lapilli) in the basal sub-unit (B7-a) at all sites (Figs. \ref{b_overall_notext} and \ref{a3_photo}a-c), scoria ashes within B7-c at all sites except for CC, and well-rounded low-density pumice within B7-d at CC and B7-c-d at WS. The established sub-unit correlations for tephras in Sequences C and B are shown in Table \ref{t_cor_b}.

Tephras in Sequence A have been observed at more than four sample sites, but the stratigraphy sometimes displays great variations between outcrops in the west and east of the Mono Craters. Their sub-unit correlations are therefore made separately.

For outcrops on the western side of the field area, the distinct feature of A4 is the presence of a non-graded, gray, sandy ash (A4-d at HMC, southwest of Mono Lake; A4-c at RC1, south-southwest of Mono Lake; central portion of A4-b at MC, northwest of Mono Lake; Fig. \ref{a4p}) overlying two to three coarse normally-graded layers (Fig. \ref{a3_photo}d and c). Therefore A4-a-b at MC, A4-a-d at HMC, and A4-a-c at RC1 are correlated with each other (Table \ref{t_cor_a}). We did not attempt to correlate A4 at ND with any other outcrops, as the pyroclast deposit was eroded and reworked.

A4 exposed at sites east of the Mono Craters is highly consistent. Constrained by the massive gray sandy ash that is also observed at DC (A4-c, southeast of Mono Lake; Fig. \ref{a3_photo}f), sub-unit correlation can be easily established for DC, WS (northwest of Mono Lake), and BC (north of Mono Lake) as shown in Table \ref{t_cor_a} and Fig. \ref{a4d}. However, A4-a at BC is not correlated with any sub-units. It is perhaps related to earlier products of A4 that are observed on the west side of the Mono Craters.  Connections between sample sites to the east and to the west of the Mono Craters include the non-graded massive gray sandy ash layer (e.g., A4-d at HMC, A4-c at RC1, A4-c at DC), and the complex-graded upper sub-unit observed at HMC (A4-e), DC (A4-f-g), BC (A4-e), and WS (A4-d).

For A3 on the west side of the Mono Craters, the presence of ostracod shells within A3-a at HMC (southwest of Mono Lake) and within A3-a-b at RC1 (south-southwest of Mono Lake) supports their correlation (Fig. \ref{a3p}). A3 at ND1 is a single tephra unit. Based on its normal grading and grain size distribution, we speculate that A3 at ND1 may be correlated with A3-c-d at RC1. The inverse grading bridges A3-b at HMC with the lower portion of A3-f at RC1.

On the east side of the Mono Craters, the presence of dark laminae and change in color between sub-units play major roles in sub-unit correlation. The established sub-unit correlation is shown in Table \ref{t_cor_a} and Fig. \ref{a3d} (Fig. \ref{a3_photo}g-i). Sample site MC (northwest of Mono Lake) is included here because of its similarity with BC (north of Mono Lake) and WS (northeast of Mono Lake). The highly consistent variation in grain size from A3-c-d at DC and A3-a-c at WS constructs their correlation. The stratigraphy of A3-a-f at both MC and BC is highly consistent, and is hence correlated (Table \ref{t_cor_a} and Fig. \ref{a3d}). Since HMC and MC are in the same direction with respect to the Mono Craters, they are likely to be correlated, and their grain size varies consistently with sub-units (from bottom to top: fine to coarse to fine). Therefore, it is inferred that A3-a-c at HMC and MC is correlated. For A2, the correlation is made based on the relative thickness of each sub-unit at different sample sites (see Fig. \ref{a2o} and Table \ref{t_cor_a} for more details).

Lake silt divides A1 into two portions at HMC, RC1, ND2, and MC, which provides the basis for correlation (Figs. \ref{a1p} and \ref{a3_photo}j-l). The lower portion of A1 at these sites is highly consistent, and is characterized by couplets of coarse and fine layers (at MC, the coarse layers are thinner, dark-gray laminae). Above the lake silt, at all sites except for WS, the most distinct feature for A1 is the presence of an ungraded, massive sub-unit intercepted by one or two thin fine lamina (A1-b at MC, HMC, A1-g at RC1, and A1-d at ND2; Fig. \ref{a1p}). A rough one-to-one correspondence for these sub-units can be established for these sites (Table \ref{t_cor_a} and Fig. \ref{a3_photo}j-l).

\subsection{Noticeable features of the WCF tephras and their interpretation}
The stratigraphy of tephras in the WCF and their notable features are highlighted in this section. Most of the WCF tephras are thicker at outcrops to the northeast of the Mono Craters, suggesting a prevailing wind blowing towards the northeast \citep{L68}. This is consistent with transport patterns of tephras from the Mono Craters during the Holocene. The exception is A1, the products of which were all blown towards the northwest.

In terms of stratigraphy and sedimentology, the most distinct feature of the tephras within the WCF is a couplet consisting of a coarse tephra overlain by a fine (silty to fine-sandy ash) tephra layer (e.g., A4, A3, and A1 as shown in Fig. \ref{a3_photo}d-l). Its main distinction from a normally-graded sub-unit is that a contact can be easily recognized between the layers. The increased difference in settling velocity between the coarse and fine tephra under water together with the waning of an eruption led to the development of separate layers with sharp contact. The frequent occurrence of couplets suggests that most of them were originated from a single non-sustained eruption pulse. This eruptive behavior is consistent with the pulsatory subplinian eruptions from the Mono Craters during the Holocene \citep{sieh1986most, bursik1993subplinian, bursik2014deposits}. Potential exceptions are the couplets whose lower coarse and upper fine layers share comparable thickness. In such cases (A3-b-c at MC; Fig. \ref{a3_photo}g), their total grain size distribution is platykurtic, and it is possible that they were produced from two eruption pulses (the latter being more highly-fragmented). Currently whether such couplets were from one or two eruption pulses is inferred based on sub-unit correlation and their thickness variability at different sites. Other processes, such as settling-driven convection \citep{hoyal1999influenceb}, ice rafting \citep{zimmerman2011freshwater}, and disintegration of aggregates, could also potentially affect the formation of tephra couplets within the WCF.

The uppermost portion of some tephra units (D17, C14, C10, C9, A4, and A3; see A4-e at HMC in Fig. \ref{a3_photo}d as an example) is characterized by numerous thin couplets (several millimeters for one couplet) of coarse and fine layers with comparable thickness. This resembles the Gray Glassy Beds (GGB) of the North Mono eruption \citep{sieh1986most}, but the GGB are rich in lithics and obsidian \citep{bursik1993subplinian}. We are not yet sure if such couplets within the WCF tephras are lithic-rich, because each of them is too thin to be sampled individually, and particles within them are too fine. However, the coarse layers within these couplets with darker color suggest that they may be obsidian-rich. We interpret them to be the products from small eruption pulses during the last stage of an eruption. It remains hard to tell if they were related to ash venting activities (e.g., \citealp{bonadonna2002tephra,cole2014ash,black2016ash}), but they might reflect the nature of the corresponding eruptions and the Mono Craters.

Tephra grains within the WCF are mostly sub-angular and denser than water. This characteristic is so pervasive that for some layers, it can be questioned whether low-density pumice (density less than water) was ever produced. Sparse low-density and rounded pumices are found within B7-d at CC and WS, A3-a at DC, and A3-d at DC.

It takes time for low-density pumices to get saturated to be able to sink and deposit if they fall on a water body \citep{whitham1986pumice,manville1998saturation,white2001settling,risso2002presence,vella2007waterlogging,fauria2017trapped}. The saturation process provides time for them to transport and get abraded as a result of particle collision from wind-induced wave. Since coarser pumice grains require longer saturation time \citep{manville1998saturation}, it is possible that a lot more coarser pumice was also erupted during the eruptions that produced B7-d, A3-a and A3-d at DC, but was transported farther while floating above water and deposited elsewhere later. This interpretation is consistent with the rounded shape of the pumices within B7-d and the fact that the low-density pumices within A3-a and A3-d at DC being the coarsest grains. If that is the case, the fine material within B7-d was then probably formed partially due to abrasion, and it is possible that the true volume of A3 and B7 might be greater than the estimate derived from the observed thickness.

Following the same logic, it is reasonable to infer that other eruptions could have produced low-density pumice as well, but it was not deposited at the lake bottom after the eruption. This can be regarded as another possible interpretation to the coarse-fine couplets that are commonly observed within the WCF tephras, as more fines were produced due to abrasion. If low-density pumice was indeed produced from B7, A3, and perhaps other eruptions associated with the WCF, the next question to ask is where the low-density pumice was finally deposited. The high-resolution record of lake level fluctuations shows that Mono Lake remained hydrologically closed more often than not and for long periods during the late Pleistocene \citep{zimmerman2011high}, yet distinct layers composed of abraded, rounded, low-density pumice have not been observed within the WCF \citep{L68,zimmerman2011high}. Further study on deposits above each WCF tephra layer could help reveal whether more low-density pumice was ever produced, and, if so, their transportation and deposition pattern.

For the WCF tephras exposed at distal sites, the deposits below and above the WCF tephras are mostly laminated or massive lake silt, suggesting a stable depositional environment. At proximal sample sites, the WCF tephras could be in fact fall, pyroclastic flow and surge deposits, or have been reworked after deposition. The fact that the observed pyroclastic or epiclastic deposits have transported both in the atmosphere and below water increases the difficulty in their interpretation. Wavy lamination and imbrication of coarser pyroclasts can be observed at RC and ND (e.g., A3-c-d at RC1; Fig. \ref{a3p}), suggesting that the lacustrine environment did affect the sedimentation process in some cases.

To identify pyroclastic flow deposits, our present criteria are based on hummocky cross-lamination, poor sorting of the silty ash to fine-sandy ash, matrix supported lapilli-sized pumice, and only proximal occurrence (correlatble at RC and ND, but absent at HMC, a site that is $2.5-4$ km from RC and ND). If the underlying deposit is lacustrine or well-sorted primary pyroclastic, the likelihood of being reworked is small given the paucity of source material. If a poorly-sorted primary pyroclastic deposit is reworked, the coarser particles are likely to concentrate at the base due to greater settling speed.

Based on these criteria, we propose upper portion of A3-f at RC1 (Fig. \ref{a3p}), and A1-a at ND2 (Figs. \ref{a1p} and \ref{a3_photo}j) as pyroclastic flow deposits. A4-e at RC1 may be a reworked pyroclastic flow deposit (Fig. \ref{a4p}), as it has all features listed above, and its base is characterized by concentrated lapilli-size pumice. The fact that almost all other sub-units (from fallout) at RC and ND can be somewhat correlated with sub-units at HMC or DC is consistent with the current interpretation. In addition, the lowermost portion of A1 at ND2 displays unique features characteristic of surge deposits (Figs. \ref{a1p} and \ref{a3_photo}j), which will be discussed in the following text.

\subsection{Interpretation of B7 and tephras in Sequence A}
\subsubsection{B7}

Sub-units of B7 are well-correlated at sample sites SEB, TS, CC, and WS, and have comparable thickness at these locations (Figs. \ref{b_overall_notext} and \ref{a3_photo}a-c). For B7-a, the mode of its grain size is $1-0$ $\phi$ at SEB, CC, and WS, and is $2-1$ $\phi$ at TS. These observations imply that the transport of B7-a was widespread, and the associate eruption is greater in volume compared to other tephra sub-units in Sequence B. Its greater thickness at TS suggests a prevailing wind blowing towards the west or northwest. Comparison with well-studied Holocene tephras from the Mono Craters with similar thickness distribution \citep{sieh1986most,nawotniak2010subplinian,bursik2014deposits} suggests that its volume is likely comparable to the Upper Beds of the South Mono eruption, the most intense and voluminous eruptive phase from the penultimate eruption from the Mono Craters (594-648 cal A.D.; \citealp{bursik2014deposits}). Therefore, we suggest that the volume of B7-a is $\sim0.1-0.5$ km$^3$.

Sub-unit B7-c at SEB and WS includes three distinct scoria layers (Fig. \ref{a3_photo}a). The scoria is mixed within B7-c at TS, and negligible amount ($<0.1\%$) of scoria is observed within B7-c at CC. The spatial distribution of these sample sites suggests that if the scoria ash was produced from the Black Point Volcano, the resultant deposit should be thick at TS, and observable at CC given its occurrence at WS (as they are in the same direction with respect to the Black Point Volcano). These features are not observed for the basaltic ash layers within B7-c. Possible basaltic source vents during the time of B7 are the Black Point Volcano and the scoria cone at June Lake, near the southern end of the Mono Craters (which produced B7* below B7; Fig. \ref{ggb}a; \citealp{bailey1989geologic,bursik1993late,hildreth04}), which makes us speculate that these scoria layers were from the latter. 

Sparse rounded low-density pumice within B7-d can be explained by its transport and deposition above the water surface and in subaqueous environment \citep{whitham1986pumice,manville1998saturation,white2001settling,risso2002presence,vella2007waterlogging,fauria2017trapped}. Reworking is another possible interpretation to B7-d. However, if that is the case, the well-saturated, heavier,  lapilli-sized pumice would settle first, and deposit at the base of B7-d \citep{manville1998saturation}. This contradicts to our field observations (the well-rounded low-density pumice is mixed within B7-d), which suggests reworking as a less likely interpretation.

We observe a distinct reworked layer above B7 at WS and possibly at TS (Fig. \ref{b_overall_notext}; Fig. \ref{a3_photo}b and c). The absence of such a layer at SEB, a sample site much closer to the Mono Craters and located in the same direction as TS with respect to the Mono Craters, suggests that it was not directly produced from the Mono Craters. We hypothesize that this may be related to lake water level fluctuation during that time, or to the presence of low-density pumice within B7-d. The occurrence of many distinct features within B7 and its surrounding deposits suggests a potential causal relationship. Its uniqueness in physical characteristics and stratigraphy are worth further investigation.

\subsubsection{A4}
Outcrops of A4 have products from at least six distinguishable eruption pulses. Tephra deposits (e.g., A4-a-c at HMC; Fig. \ref{a3_photo}d) from the first three are similar in stratigraphy and grain size distribution. The sub-unit correlation shows that dispersal was mainly towards the west or northwest. These deposits are lithic-poor, and the gradual change in grain size between sub-units observed at RC1 suggest that intervals between these eruptions were short. No deposit from the third eruption pulse occurs at RC1 possibly due to erosion. The overlying sub-unit of A4 (A4-d at HMC and A4-c at DC) is a massive, well-sorted, non-graded, gray medium-sandy ash. It was transported over a large area, covering HMC, RC1, DC, and perhaps MC and BC. These features indicate that there might have been a sudden change in eruption style or other eruption parameters.

The tephra from the next eruption pulse was transported northwards. Limited reworked deposit (?) was preserved at RC1 (A4-d at RC1). It is rhyolitic lithic-rich at DC (A4-e at DC; Fig. \ref{a4d}), possibly suggesting that a new vent was opened, and accompanied by the fragmentation of an existing dome. The last stage of the A4 eruption produced complex graded products (fine to coarse to fine) deposited at HMC, DC, BC, and WS as A4-e (Fig. \ref{a3_photo}d), A4-f-g (Fig. \ref{a3_photo}f), A4-e, and A4-d, respectively (Figs. \ref{a4p} and \ref{a4d}). These suggest that there might have been two or three eruption pulses, and that their products were transported over a great area, and were not strongly directed by wind. At RC1, sub-units with similar features are not found due to erosion. The uppermost sub-unit of A4 at RC1, A4-e, is probably a reworked dilute pyroclastic flow deposit (thick, massive, poorly-sorted, and the underlying deposit was eroded). It was formed after the deposition of all observed A4 tephra sub-units.

Load cast structures, and convoluted and contorted laminations have been observed at the uppermost sub-units of HMC (A4-e), RC1 (A4-e), and ND2 (A4-c), respectively (marked in Fig. \ref{a4p}). These structures can be formed due to liquefaction of saturated sediment during seismic events. Cross-cutting relationship suggests that they were all formed after the deposition of A4 and prior to the eruptions that produced A3. The lake silt between A4 and A3 is not thick (thickness: 3.2, 10.2, and 0 cm at HMC, RC1, and ND2, respectively) at these sites. Based on these observations, we speculate that the uppermost sub-unit of A4 recorded a seismic event after its deposition. Since the basal sub-units of A3 have sharp and planar lower contact (except for ND2), we further infer that the seismic event might mark the onset of the A3-associated eruptions. However, we stress that more work is required to confirm these hypotheses.

\subsubsection{A3}
\begin{figure}[H]
	{\includegraphics[width=\textwidth]{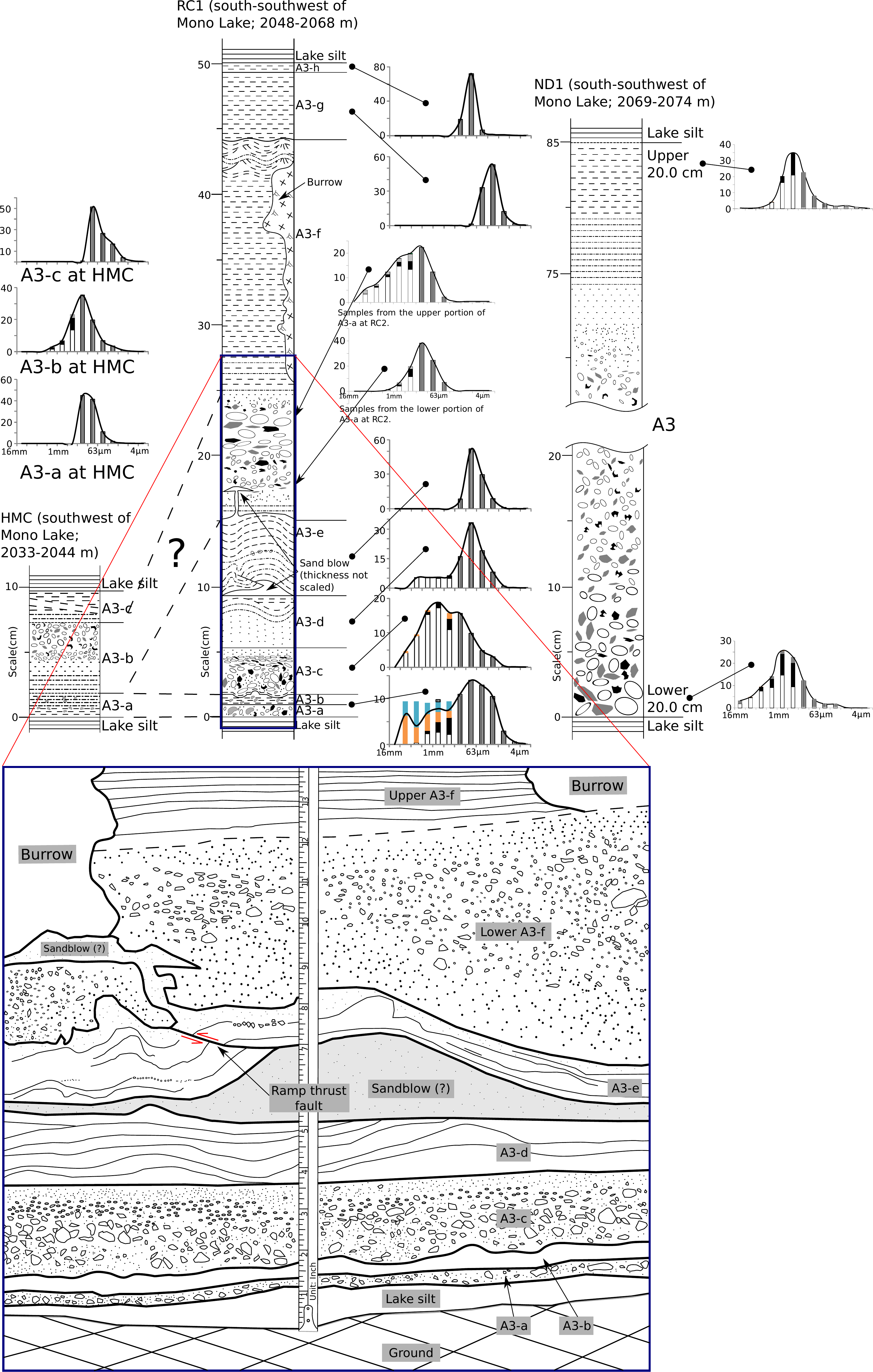}}
	\caption{Stratigraphic columns of A3 at HMC, RC1, ND1, and grain size distributions of sampled tephra. These outcrops are plotted together as they are proximal to the Mono Craters, and can be well-correlated. Proposed sub-unit correlation is marked with dashed lines. See Fig. \ref{d_overall} for explanation of symbols. The elevation range of the observed tephras at each sample site (as shown in Fig. \ref{ggb}b) is given in bracket. }
	\label{a3p}
\end{figure}

\begin{figure}[H]
	{\includegraphics[width=\textwidth]{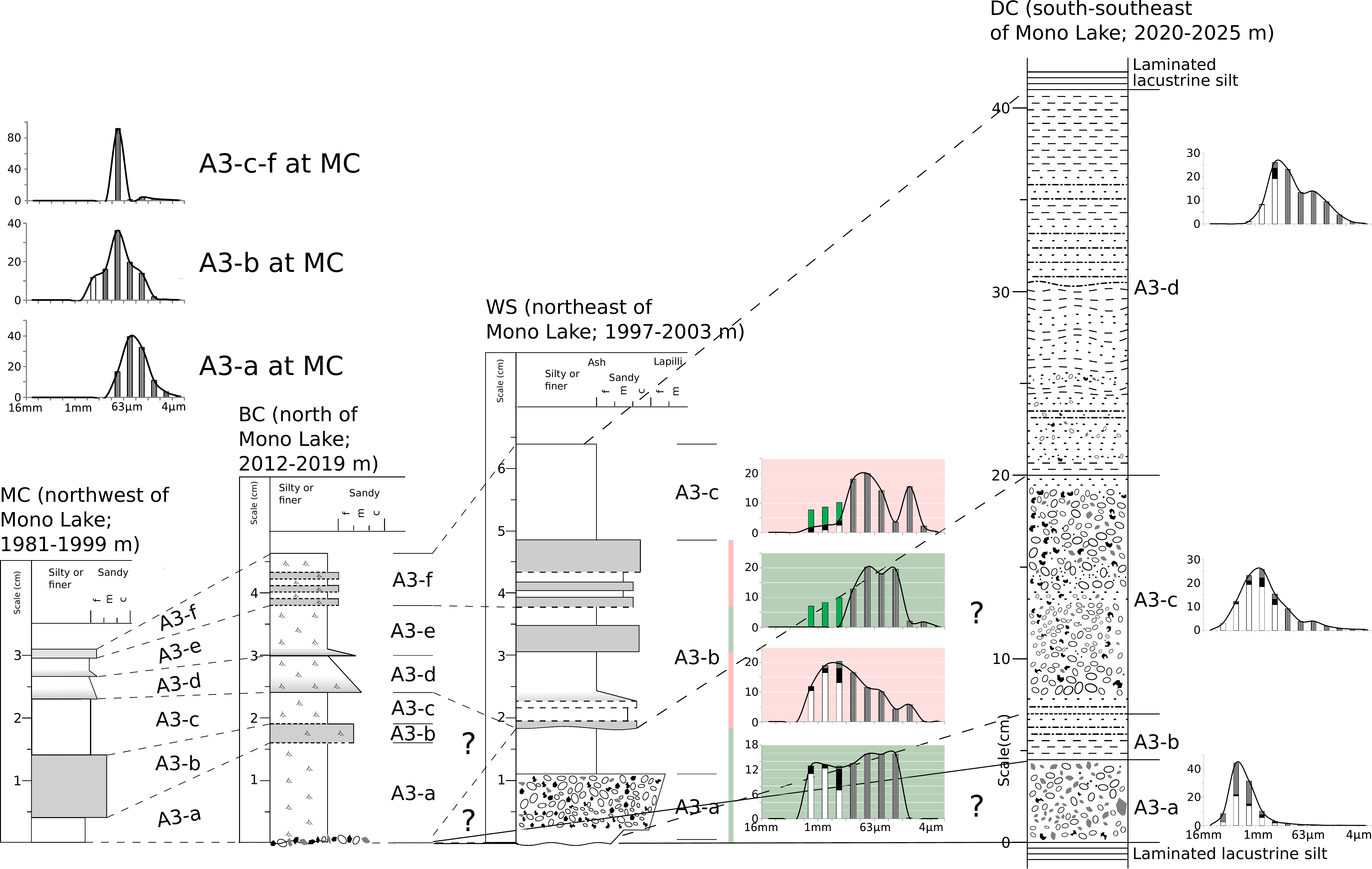}}
	\caption{Stratigraphic columns of A3 at MC, BC, WS, DC, and grain size distributions of sampled tephra. MC is included in this figure as it is well-correlated with A3 at BC. Proposed sub-unit correlation is marked with dashed lines. The scale of DC is different from that of MC, BC, and WS. Note that A3 at WS was not sampled according to sub-units, and color bars on the right of the stratigraphic column show the sampled stratigraphic positions, which correspond to the colors of the grain size distribution charts on the right. See Figs. \ref{d_overall} and \ref{a4p} for explanation of symbols for A3 at DC and MC, BC, and WS, respectively. The elevation range of the observed tephras at each sample site (as shown in Fig. \ref{ggb}b) is given in bracket. }
	\label{a3d}
\end{figure}

\begin{figure}[H]
	{\includegraphics[width=\textwidth]{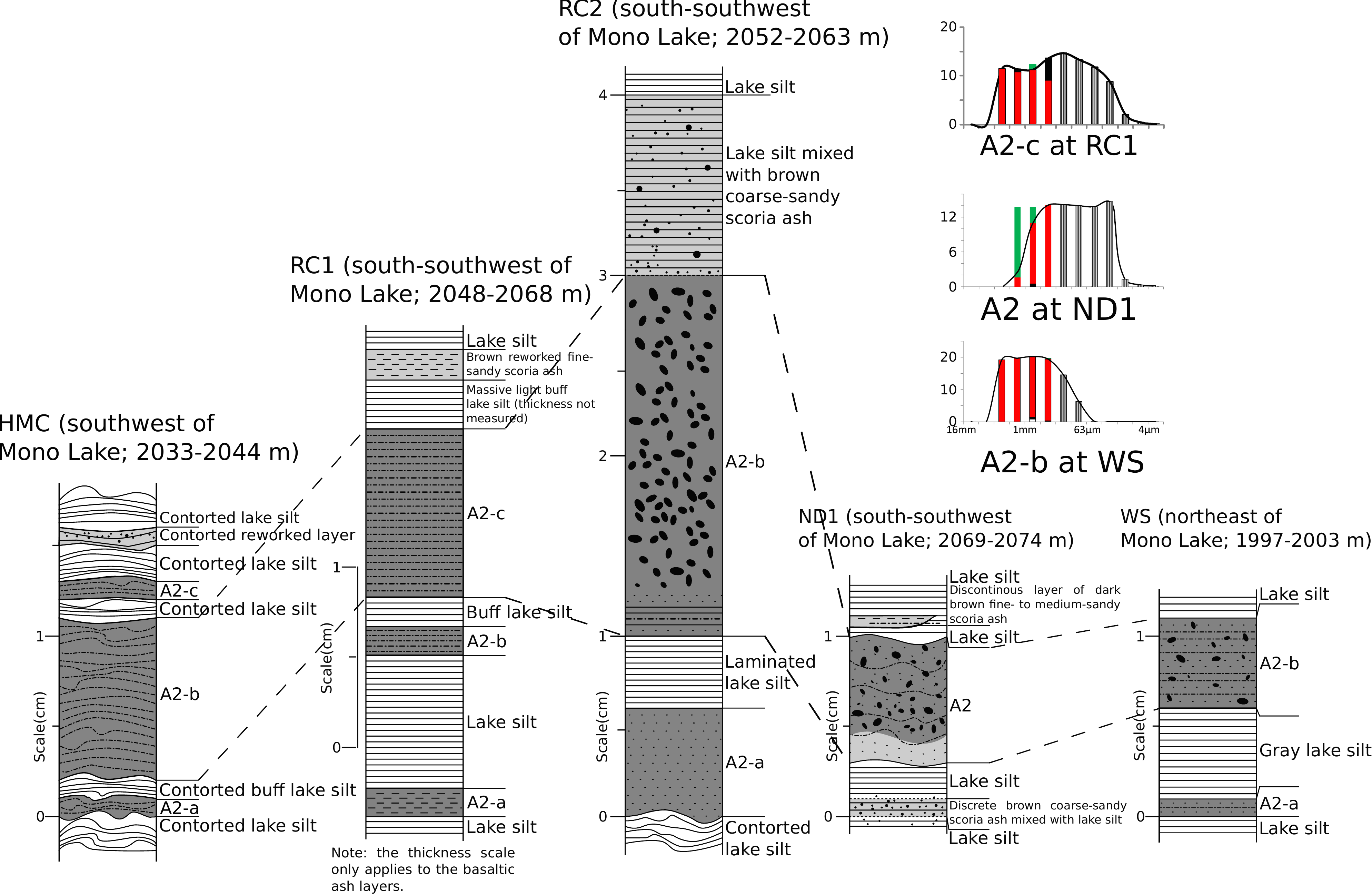}}
	\caption{Stratigraphic columns of A2 at HMC, RC1, RC2, ND1, WS, and grain size distributions of sampled tephra sub-units. Proposed sub-unit correlation is marked with dashed lines. See Fig. \ref{d_overall} for explanation of symbols. The elevation range of the observed tephras at each sample site (as shown in Fig. \ref{ggb}b) is given in bracket. }
	\label{a2o}
\end{figure}

\begin{figure}[H]
	{\includegraphics[width=\textwidth]{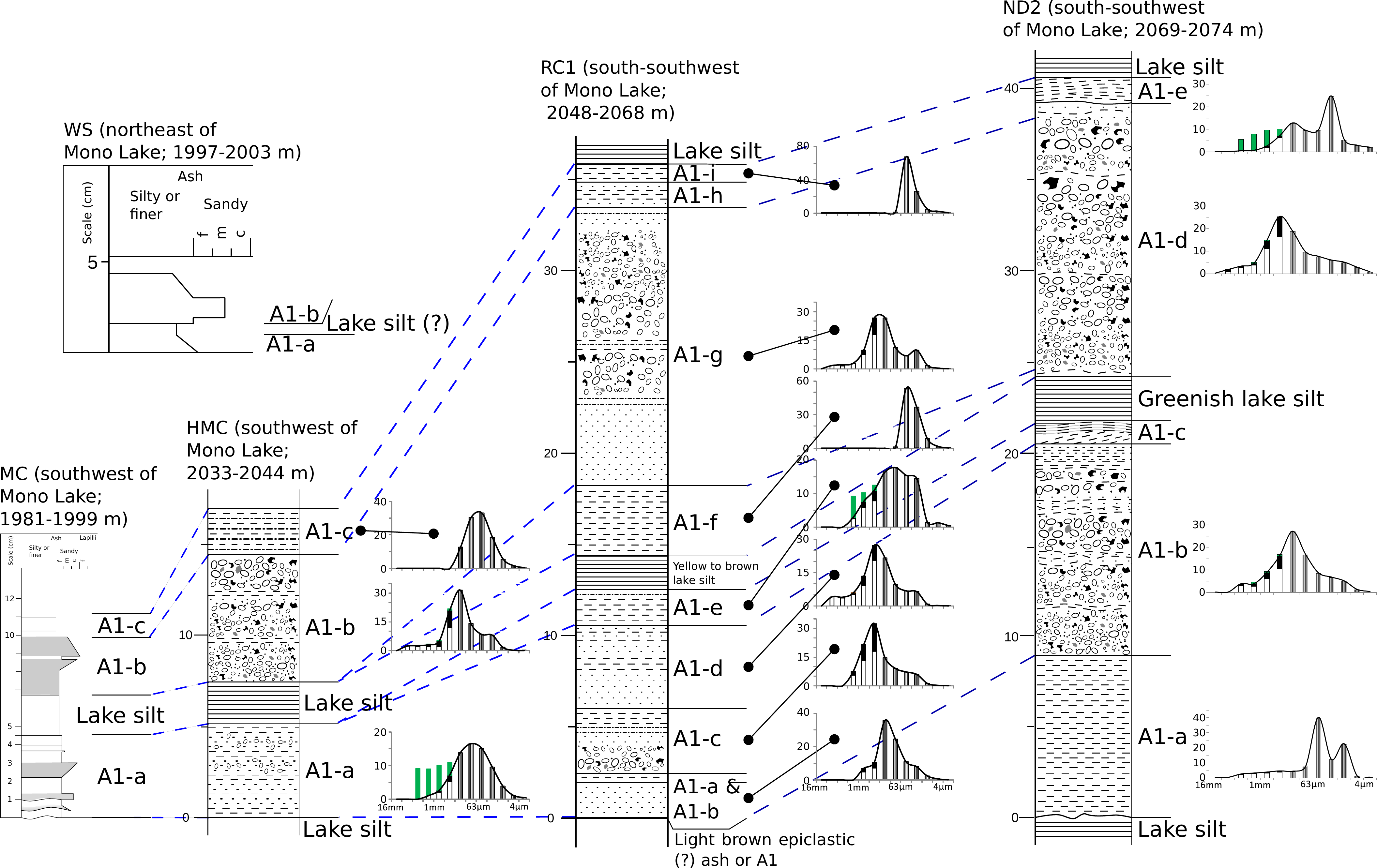}}
	\caption{Stratigraphic columns of A1 at MC, HMC, RC1, ND2, WS, and grain size distributions of sampled tephra. Proposed sub-unit correlation is marked with dashed lines. See Figs. \ref{d_overall} and \ref{a4p} for explanation of symbols for A1 at HMC, RC1, and ND2, and WS and MC, respectively. The elevation range of the observed tephras at each sample site (as shown in Fig. \ref{ggb}b) is given in bracket. }
	\label{a1p}
\end{figure}

\begin{figure}[H]
	{\includegraphics[width=\textwidth]{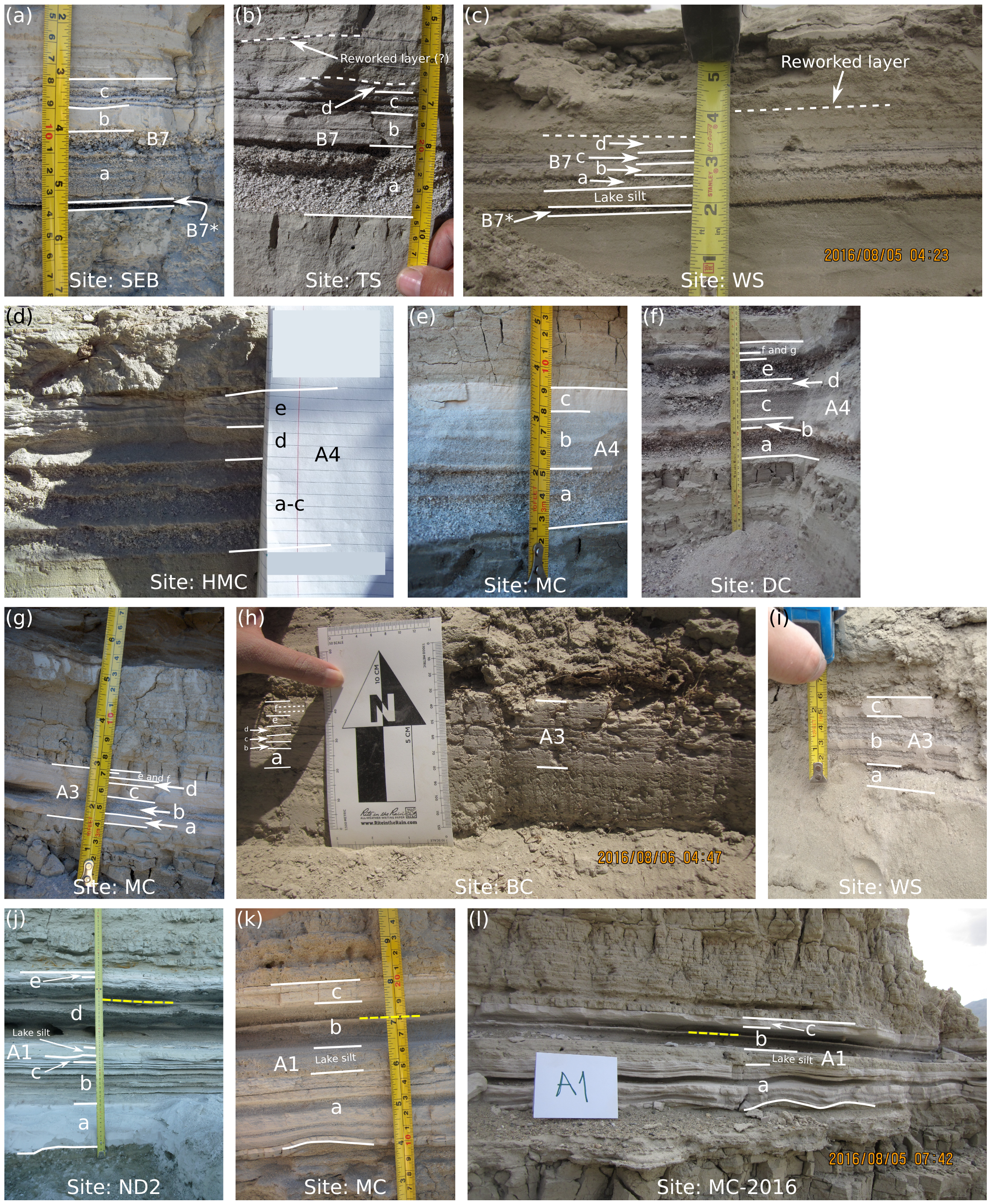}}
	\caption{Selected pictures of outcrops. a-c: outcrops of B7 (and B7*) at SEB, TS, and WS; d-f: outcrops of A4 at HMC, MC, and DC; g-i: outcrops of A3 at MC, BC, and WS; j-l: outcrops of A1 at ND2, MC, and MC-2016. Their sub-units are labeled. In g-i, we use yellow dashed lines to highlight a thin lamina of silty ash observed at the three sites. Note the wavy lower contact and wavy lamination of A1-a in h and i.}
	\label{a3_photo}
\end{figure}

\begin{figure}[H]
	{\includegraphics[width=0.7\textwidth]{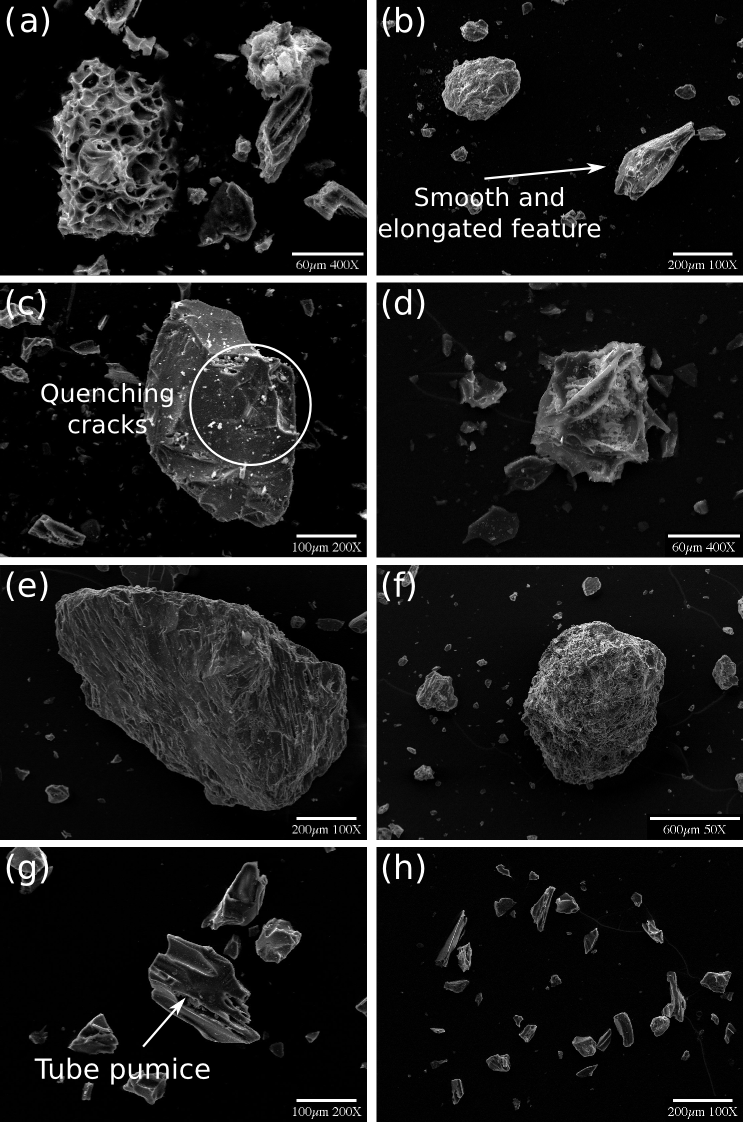}}
	\caption{Selected SEM images: a: high-porosity tephra from E18; b: tephra grains with smooth and elongated feature from C12 at SEB; c and d: samples from C10* at SEB (with quenching cracks in c); e and f: blocky tube pumice and aggregate (?) of B7-a from SEB; g and h: tephra grains of A3 from ND1 and A3-h from RC1, which are  mainly tube pumices and glass shards.}
	\label{sem}
\end{figure}

Outcrops of A3 contain three or four well-defined beds (at HMC, these are A3-a-c; at RC1 these are A3-a-b, A3-c-d, and the lower portion of A3-f; Fig. \ref{a3p}), plus at least five additional much thinner beds (at DC and BC, they are in A3-d and A3-d-f, respectively; see Fig. \ref{a3_photo}h for outcrop of A3-d-f at BC) as later products of the A3 eruptions.

Earliest products of A3 include the correlated A3-a at HMC and A3-a-b at RC1 (Fig. \ref{a3p}). This thin layer is only found at the two sites, and is likely small in volume. A3-a-b at RC1 are rich in fragments of welded Bishop Tuff and ostracods. The presence of ostracods indicates that A3-a-b at RC1 mark the early stage of the A3 eruption, which excavated ostracod-bearing lake bed sediments, but did not disrupt older domes (limited amount of rhyolitic lithics). Together with the presence of Bishop Tuff, it is inferred that A3-a-b at RC1 were derived from a variety of depths, which is consistent with their lowermost stratigraphic position within A3 \citep{bursik1993subplinian}. Another early eruption pulse produced lithic-rich tephra deposited at DC (A3-a-b). The abundant rhyolitic lithics within A3-a at DC (Fig. \ref{a3d}) suggests that the corresponding eruption pulse did erode existing domes, and it is more likely that A3-a-b at RC1 predates A3-a-b at DC.

The next eruption pulse deposited tephra at RC1 (A3-c-d), DC (A3-c), and WS (A3-a; Fig. \ref{a3_photo}i), which implies a dispersal axis towards the northeast. The sub-units are normally graded at RC1, but the imbrication and wavy lamination at RC1 suggest that its sedimentation there was affected by the lacustrine environment (Fig. \ref{a3p}). At DC, the central portion of this sub-unit (A3-c) is rich in fine material compared to its lowermost and uppermost portions, implying that there was a decrease in column height or change in wind direction during the eruption.

Following this eruptive pulse, occurrence of an earthquake is temporarily proposed by the contorted cone-shaped deposit (known as sand blow) observed at RC1 (lowermost portion of A3-e; Fig. \ref{a3p}). After the earthquake, a few eruptive pulses of small volume left deposits only observed at RC1 (A3-e; Fig. \ref{a3p}). Judging by their differences in color and grain size, and well-defined contacts,  they were produced from three distinct pulses. Another earthquake may have taken place after their deposition, as indicated by the ramp-thrust fault cutting across the lower portion of A3-e at RC1 (Fig. \ref{a3p}). This earthquake may have marked the onset of the subsequent eruption, producing lower A3-f at RC1, characterized by inverse grading. This thick sub-unit (A3-f) is well-correlated among sites RC1, RC2, HMC, and MC (Fig. \ref{a3p}), suggesting that the tephra was dispersed northwestwards. Its inverse grading suggests that the column height of the eruption progressively increased, and the increase in rhyolitic lithic content upwards (based on the grain size distribution of A3-a at RC2 shown in Fig. \ref{a3p}) is possibly indicative of vent widening during the eruption. A pyroclastic flow probably occurred after the sustained eruption, as suggested by the hummocky cross-stratification of the upper portion of A3-f at RC1 (Fig. \ref{a3p}). There is no well-defined contact between the lower (inversely-graded fall deposit) and upper (hummocky cross-stratification) portions of A3-f (Fig. \ref{a3p}), suggesting that the latter was emplaced soon after the sustained eruption (which produced the lower portion of A3-f). It is possible that there is a causal relationship between the two types of deposits within A3-f.

Upper deposits of A3 were all transported to the north or northeast. They are composed of at least five couplets of coarse and fine layers with varying thickness (Fig. \ref{a3d} and see A3 at BC and WS as examples in Fig. \ref{a3_photo}h and i). Their relative thicknesses suggest that the lower two might be greater in volume. In addition, upper sub-units of A3 at RC1 seem to record two distinct tephra fall deposits, but no correlation can be made given current observations.

\subsubsection{A2}
Detailed work has previously been done on the stratigraphy and pyroclastic characteristics of the deposits of the Black Point Volcano \citep{custer1973stratigraphy,white2000subaqueous,murtagh2013pyroclast}.  \cite{murtagh2013pyroclast} suggest that the eruption might have been of violent Strombolian or Plinian style based on the study of pyroclast textures. It is worth noting in this contribution that the stratigraphy of A2 suggests that there were at least three eruptive pulses, of which the products reached at least 28 km from source, with the second pulse being greatest in volume. Thickness measurements suggest a prevailing wind blowing towards the east-southeast for the second, main eruptive pulse.

We use the trend model of \cite{yang2016} to construct the isopach map of the tephra deposit from the second eruption pulse of A2, and methods proposed by \cite{pyle1989thickness} and \cite{fierstein1992another} to estimate its volume. Exponential decay in thickness with the square root of isopach area is assumed for simplicity. By trying different values in wind direction to minimize the sum of squared residuals, it is inferred that the prevailing wind direction ranged from $110-130^\circ$.  The estimated volume ranges from $0.11$ to $0.25$ km$^3$. This value is slightly higher than the previous estimate of $0.088$ km$^3$ for the Black Point ``tephra sheet'' \citep{custer1973stratigraphy}, obtained by adding areas multiplied by mean thicknesses within isopachs by hand.

Previous work \citep{murtagh2013pyroclast} on proximal deposits of A2 suggest a violent eruption. Thickness of A2 at all sample sites suggests a deposit with great extent and low thinning rate, supporting the inference that sub-Plinian to Plinian eruptions took place. In addition, deposits from the second eruption pulse of A2 are poorly-sorted, and lack grading (Fig. \ref{a2o}), suggesting that particles with a wide range of grain size were produced from the eruption, and their deposition was not strongly controlled by grain size or waning of the eruption. This is consistent with the inference that both magmatic and phreatomagmatic styles were involved in the eruption \citep{murtagh2013pyroclast}.

\subsubsection{A1}
Sub-units within A1 can be correlated between all sites except for WS (Figs. \ref{a1p} and \ref{a3_photo}j-l), and the interbedded lake silt (thickness: 2.2, 1.9, 2.4, 1.7 cm at HMC, RC1, ND2, and MC, respectively) within it suggests a long hiatus during its eruption. The earliest product (A1-a at ND2) of the A1 eruptions is found only at RC1 and ND2, and recognized as dilute pyroclastic flow deposit (Fig. \ref{a3_photo}j). Later products of the A1 eruption before the long break are composed of more than six couplets of coarse (non-graded coarse-sandy ash to fine lapilli) and fine layers (massive silty ash). There are six of them at sites RC1, HMC, and MC (Fig. \ref{a3_photo}k and l), while ND2 preserves more than ten couplets (Fig. \ref{a3_photo}j) that are also coarser than the ones observed elsewhere.

The number of couplets at RC1 and ND2 does not match -- i.e., at two sites that are $<2.5$ km from each other -- implying that the preservation of additional fine layers at ND2 is related to its proximity to the source vent. It is thus possible that not all fine layers at ND2 are tephra fall deposits, i.e., deposits that blanket the terrain.  We suggest that some fine layers within the couplets of A1-b at ND2 are surge deposits. This combination of features resembles the fall and surge deposits within the Orange-Brown Beds from the South Mono eruption, the penultimate eruption from the Mono Craters \citep{bursik2014deposits}. Together with the higher fraction of obsidian within these couplets (Fig. \ref{a1p}), it is further inferred that these surge deposits were probably formed as a result of water-magma interaction, which dominated the eruption prior to the deposition of lake silt within A1.

In addition to the couplets, there are one or two well-defined, massive tephra sub-units of silty to fine-sandy ash deposited prior to the significant break of the A1 eruption (Figs. \ref{a1p} and \ref{a3_photo}j-l). They are only observed at RC1 and ND2, and may have been reworked. Deposits above the lake sediment include one or two tephra sub-units that are only observed at RC1 (lower portion of A1-g at RC1), fallout deposits from a sustained eruption pulse (A1-b at HMC, Upper portion of A1-g at RC1, and A1-d at ND2) and a few more short-term ones (A1-c at HMC, A1-h-i at RC1, and A1-e at ND1). The sub-unit correlation (Table \ref{t_cor_a} and Fig. \ref{a1p}) indicates that the sustained eruption was interrupted by one or two short breaks, fluctuation of column height, or a change in wind direction, as suggested by the distinct, thin light-colored laminae of silty to fine-sandy ash observed at ND2 and MC (marked as yellow lines in Fig. \ref{a3_photo}j-l). The correlation suggests that during the sustained eruption, wind was blowing towards northwest. Later products of A1 are only observed and well-correlated at RC1 (A1-h-i) and ND2 (A1-e). They are tephra deposits from the last stage of the A1 eruptions. It is noticed that after the long break, the pulsatory behavior of A1 became much less apparent, and that all products from A1 were directed northwestward. The interpretation of tephras in Sequence A as well as other tephras in the WCF is summarized in Fig. \ref{summary_wcf}.

\subsection{Eruption style}
The most recent volcanic activity of the Mono Craters involved both magmatic and phreatomagmatic eruptions of subplinian size \citep{sieh1986most,bursik2014deposits}. There are two potential sources of water for phreatomagmatism: ancestral Lake Russell, and aquifers underneath the craters.  \citet{bursik1993subplinian} presented evidence that the Pleistocene gravels under the northern Mono Craters may have acted as an important aquifer, providing water to drive phreatomagmatism in the most recent Mono eruption.  Lake Russell has covered the northern part of the Mono Craters multiple times during the late Quaternary ($>2100$ m a.s.l; Fig. \ref{ggb}; see Fig. 19 of \citealt{L68} or its modified version, Fig. 3 of \citealt{zimmerman2011high}, for more details). The location of some craters and domes, which are not correlated with younger, magmatic eruptions, suggests that the corresponding vents could be under water during the Wilson Creek time (also given the extent of the WCF deposit as shown in Fig. \ref{ggb}a), indicating that magma-lake interactions could have taken place.

For the WCF tephras, concentrations of ostracods occur in some of the units or sub-units; this could happen upon mixing of rising magma with the water column or ostracod-rich, lake-bed sediments. Many tephra sub-units in the WCF are also characterized by a high fraction of obsidian and rhyolitic lithics, and relatively denser pumice (denser than water).  These observations suggest that many sub-units of tephras in the WCF are related to rapid quenching of magma or disruption of existing domes, and hence, are associated with water-magma interaction during eruption. Wet, base surge eruptions are also diagnostic of magma-water interaction.  We suggest therefore that phreatomagmatic eruptions took place during the eruptions of D17, A3, and A1, as indicated by the presence of ostracods, higher fractions of obsidian and lithics, and surge deposits (only for A1). Other sub-layers that were probably resulted from phreatomagmatism are listed in Fig. \ref{summary_wcf} (based on higher fraction of obsidian and rhyolitic lithics).

\subsection{Temporal variation of volcanic activities from the Mono Craters during the late Pleistocene}

Each rhyolitic tephra unit within the WCF is composed of several tephra sub-units, which indicates that eruptions from the Mono Craters during the late Pleistocene were pulsatory, as is typical of the more recent and well-studied Holocene eruptions \citep{bursik1993subplinian}. Ages of the WCF tephras and their varying stratigraphy suggest that eruptions from the Mono Craters during the late Pleistocene were not consistent in frequency and volume (Fig. \ref{thi_age}; \citealp{bevilacqua2018late}).

Older eruptions that produced E19-D16 have clustered ages (Fig. \ref{thi_age}), but the thickness of these deposits at the sample sites varies greatly. After a long period of quiescence ($ > \sim 15000$ years) following the eruption of D16, volcanic activity became more frequent, and produced more pulses with explosive products of greater volume, as indicated by increased deposit thickness in Sequence C (Fig. \ref{thi_age}). C15-a at SEB is a massive sub-unit (thickness: 31.0 cm) produced from a single eruption pulse. C11 is correlatable at SEB and TS, and is thicker and coarser compared to most WCF tephra sub-units observed at these two sites. It is therefore suggested that C15 and C11 are much greater in volume compared to most tephra sub-units within the WCF.

Additionally, the thin sub-units within C13 and C9 can be correlated between SEB and TS, and have comparable thickness at both sites. This suggests that the corresponding eruption pulses were not strongly affected by wind, which is common for more powerful eruption pulses with greater column height. Their thinner thickness suggests either that the powerful pulses were short-lived, or that their vents may be located in the southern Mono Craters. Except for these two tephra units, the overall greater thickness (Fig. \ref{thi_age}) and the preservation of more tephra sub-units (compared to tephras in Sequences D and B) in the rest of Sequence C imply that most of the source vents are located in the northern part of the Mono Craters during the Wilson Creek time.

From $\sim30-20$ ka, there were only three major volcanic events, which produced tephras in Sequence B.  Products from these eruptions are generally thinner than those in Sequence C, and can be divided into fewer sub-units.  Sub-unit correlation indicates that the transportation of B6 and B5 was subject to wind. We therefore suggest that B6 and B5 are smaller in volume than most Sequence C eruptions, or that they vented from the southern Mono Craters. Although B7 is also thinner at all sample sites, its sub-units are readily correlated (Fig. \ref{b_overall_notext} and Table \ref{t_cor_b}), suggesting that each sub-unit is relatively widely and uniformly dispersed, perhaps as a result of high volume and a high plume. Similar to the interpretation of C13 and C9, this combination of features also suggests that its vent may be located in the southern part of the Mono Craters or that the associated eruption pulses were short-lived. The presence of low-density pumice within the uppermost sub-units of B7 suggests a distinct degassing history towards the tail-end of these eruptions, perhaps related to a switch to more magmatic behavior. Although more than one interpretation can be used to explain the thinness of the Sequence B tephras, the presence of fewer sub-units within them suggests that the vents are located in the southern Mono Craters.

After a hiatus of $\sim9000$ years following the eruption of B5, volcanic activity from the Mono Craters reactivated. Major characteristics of the Sequence A tephras (except for A2) are similar to those of the tephras in Sequence C, i.e., the units are thick and have complex internal stratigraphy. The well-established sub-unit correlation at proximal sites allows us to analyze azimuthal variations in thickness and grain size, which suggest that the deposition of tephra at these sites was sensitive to the prevailing wind direction. The overall greater thickness of A4, A3 and A1, the presence of more sub-units within them, and the possible occurrence of pyroclastic flow and surge deposits, all imply that the vents are located near the northern end of the Mono Craters.

The stratigraphy suggests that tephras within each sequence do not only have clustered ages, but also have comparable thickness and similar stratigraphic characteristics, which implies consistent eruptive behavior for eruptions within the same sequence. Our interpretation of the source vent location for Sequences C-A tephras indicates that vent location migrated from the north (Sequence C) to the south (Sequence B), and then back to the north (Sequence A) of the Mono Craters.

\subsection{Water level fluctuation inferred from the stratigraphy of the WCF tephras}
The presence of lake silt and sharp boundaries between lacustrine and tephra deposits within the WCF was thought to suggest dominantly low-energy depositional environment  \citep{L68,Zimmerman06}, but the present stratigraphy of B7 (Figs. \ref{b_overall_notext} and \ref{a3_photo}a-c) and A1 (Figs. \ref{a1p} and \ref{a3_photo}j-l) displays contradictory observations.

B7 has been observed and well-correlated at southeast (SEB), northeast (WS), north (CC), and northwest (TS) of Mono Lake. Tephra layer B7*, the basaltic ash layer below B7, was not found in the northwest (TS) and north (CC) of Mono Lake. The lower contact of B7 at these two sites are irregular and gradual, respectively. Lacustrine deposit is absent between B7* and B7 in the southeast (SEB) of Mono Lake (Fig. \ref{a3_photo}a), but is 1.0-cm thick in the northwest (WS) of Mono Lake (Figs. \ref{b_overall_notext} and \ref{a3_photo}c). Additionally, basaltic layers within B7-c at SEB are wavy, and fine lapilli-sized scoria is well-mixed with rhyolitic ash within B7-c at TS (Figs. \ref{b_overall_notext} and \ref{a3_photo}a-b). Geological records and inferred salinity suggest lowering of lake level during the eruption of Sequence B  \citep{L68}, and that the lake level was $\sim 2030-2050$ m at the time B7 was deposited \citep{L68}, which is above all current sample sites. Given that B7 is sandwiched by lacustrine silt, the distinct stratigraphy below B7 at different sites cannot be explained by the gradual change in water level during that time. It remains hard to explain that at WS, the sample site with the greatest elevation ($\sim$1997-2003 m a.s.l.) among distal sample sites, lacustrine silt is preserved betwen B7* and B7 while being absent at SEB ($\sim$1976-1984 m a.s.l.). The stratigraphy of B7 at the present sample sites suggests an unstable depositional environment, but more work is required for a comprehensive understanding on the water level history and depositional environment of Lake Russell during the eruption of B7.

A1 at MC has a wavy lower contact with the underlying brownish fine sand (Fig. \ref{a3_photo}k). Wavy lamination is also observed at MC-2016 (Fig. \ref{a3_photo}l) for A1-a, and partly at ND2 (fine layers within A1-b). These observations suggest that in the northwest of Mono Lake (near site MC), A1 was deposited in a shallow-water, wave-dominated environment, instead of in deep and still water as previously envisaged \citep{L68}.

Water level fluctuation inferred by \cite{L68,benson1998correlation} suggests that one highstand of Lake Russell during the late Pleistocene took place after the eruption of the Black Point Volcano that produced the basaltic A2 tephra layer (age: $13610 ^{14}$C yr B.P.; \citealp{benson1998correlation}), and reached $2155$ m a.s.l. A minor drop in lake level is tentatively proposed based on the elevation of the palagonitized portion of the Black Point Volcano \citep{L68}, but was not investigated in detail by later works. Given the short interval between the eruption of A2 and A1 (age of A1: $12920^{14}$C yr B.P.; \citealp{benson1998correlation}), we think that the wavy lamination observed at A1-a at MC and MC-2016 (northwest of Mono Lake) is further supporting evidence that water level dropped abruptly during this short time period. The deposition of lake silt during the significant break of the A1 eruption and the fact that the upper portion of A1 (deposited after the significant break) is planar-laminated or massive (without reworking) indicate that the lacustrine depositional environment became more stable, alluding to elevated lake level.

Our observations on B7 and A1 are in general consistent with previous studies on water level fluctuation of Lake Russell during the late Pleistocene \citep{L68, benson1998correlation, zimmerman2011high}. The short and well-constrained time intervals during and between the deposition of B7* and B7 and during the deposition of A1 suggest that the change in water level might be more rapid than previously thought.

\begin{figure}[H]
	{\includegraphics[width=\textwidth]{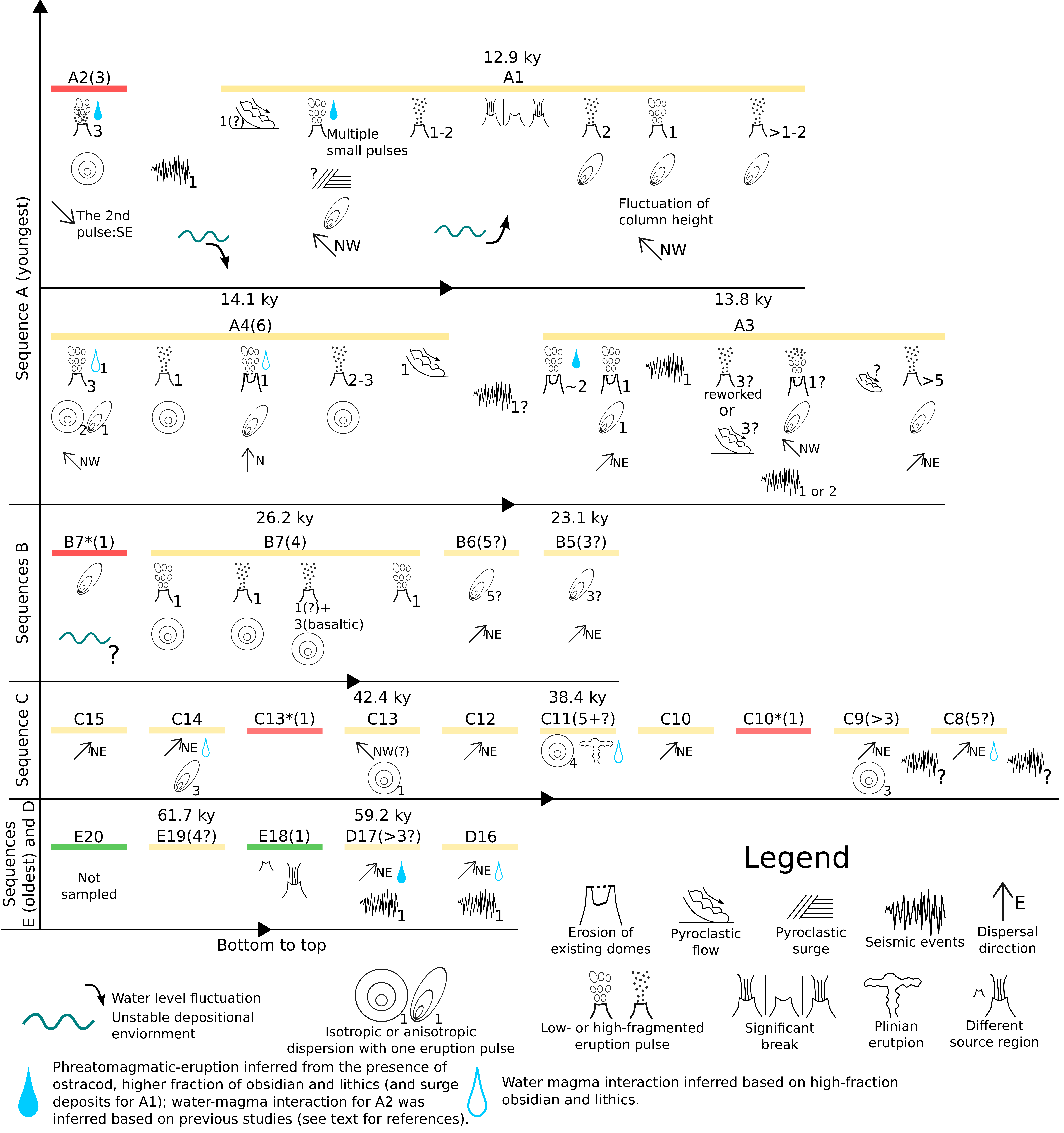}}
	\caption{Interpretation on the eruptive history of the tephras in the WCF based on current outcrops. The number of eruption pulses for each tephra unit is labeled if it can be inferred. Horizontal color bars below each tephra unit denote their compositions or origins: Red: basaltic tephras; green: E18 and E20 (rhyolitic to rhyodacitic; minor reworking) derived from Mammoth Mountain. See \cite{M14} for detailed characterization of their geochemical properties; yellow: rhyolitic. Ages of E19, D17, and C11 are referenced from \cite{vazquez2012high}, C13 from \cite{bevilacqua2018late}, B7 from \cite{M14,marcaida2015thesis,marcaida2015resolving}, B5 from \cite{chen1996edge}, and A4, A3, and A1 from \cite{benson1998correlation}.}
	\label{summary_wcf}
\end{figure}

\begin{figure}[H]
	{\includegraphics[width=\textwidth]{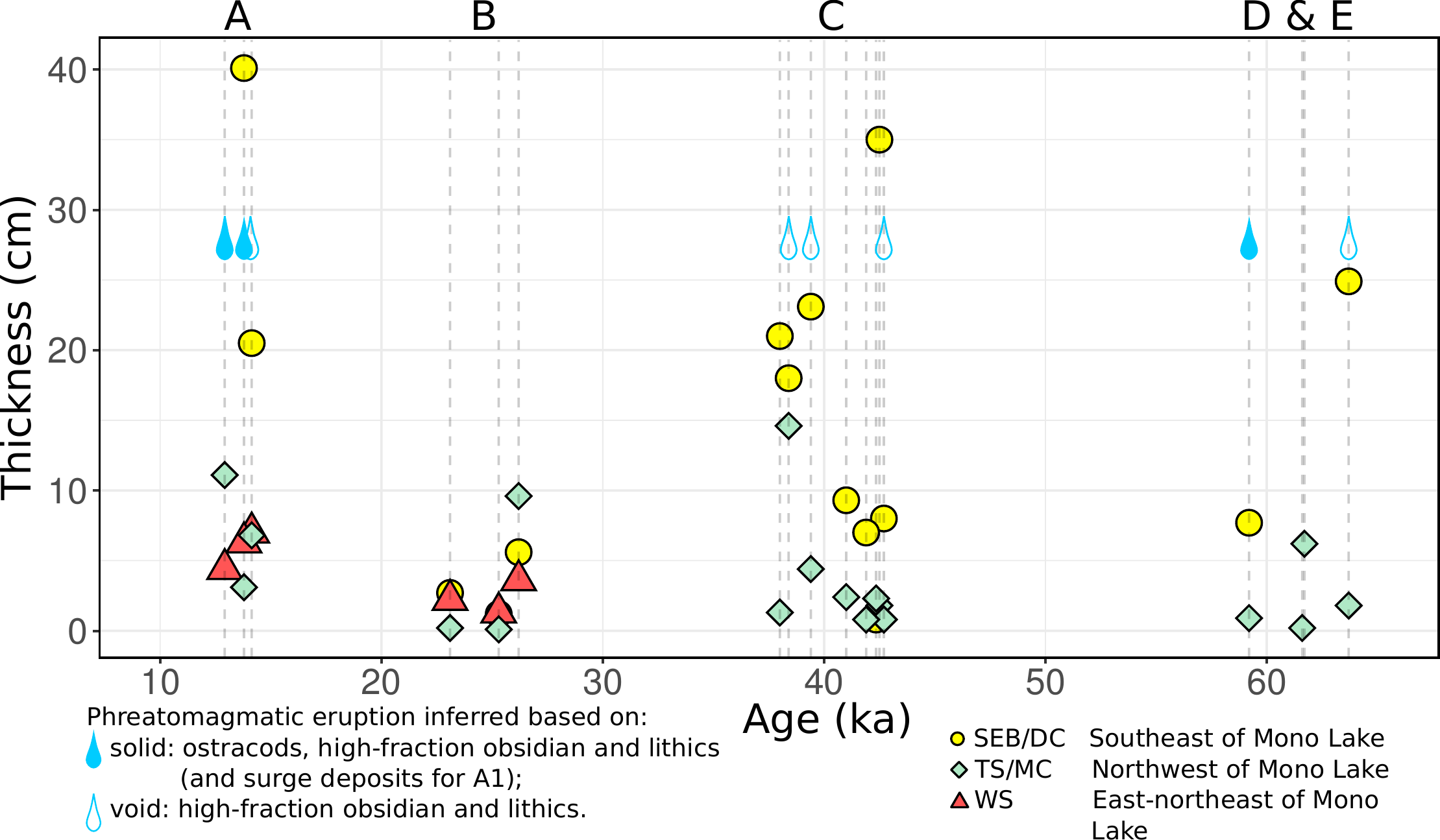}}
	\caption{Total thickness of tephras in Sequences E-A at SEB/DC (circle), TS/MC (diamond), and WS (triangle) plotted against their age. Ages of E19, D17, and C11 take reference from \cite{vazquez2012high}; D16, C14, C13, C12, C9, and C8 from \cite{bevilacqua2018late}; C15, C10, and B7 from \cite{M14,marcaida2015thesis,marcaida2015resolving} based on correlation with Domes 19, 24, and 31, respectively; B5 from \cite{chen1996edge}; A4, A3, and A1 from \cite{benson1998correlation}. A thorough database of these ages and their uncertainty is reported in \cite{bevilacqua2018late}. Ages of B6 and E18 are inferred by interpolation based on lake sediment thickness and the age of tephras below and above them. C13*, C10*, B7*, and A2 (also due to its extreme thickness at MC) are excluded here since they are basaltic, and were not derived from the Mono Craters. Whether water-magma interaction was involved during these eruptions is also marked consistently as shown in Fig. \ref{summary_wcf}.}
	\label{thi_age}
\end{figure}

Lake Russell experienced two major highstands during Wilson Creek time, whick took place prior to and after deposition of Sequence B \citep{L68}. The extent of the WCF deposits (Fig. \ref{ggb}a) suggests that the northern end of the Mono Craters was once covered by lake water during active phases of volcanism. This is consistent with our inferred vent locations for A4, A3, and A1, and the interpretation that water magma interaction was involved in these eruptions. It is worth noting that the ages of older, obsidian-rich units (D17, D16, C14, C11, C8; Fig. \ref{summary_wcf}) also coincide with the highstand of Lake Russell prior to the deposition of Sequence B. Based on our interpretation of the eruptive style (Fig. \ref{summary_wcf}), phreatomagmatism was more common during highstands of Lake Russell.

\section{Conclusions}
Previous studies have attempted to reconstruct the eruptive history of the Mono-Inyo Craters based on inferences from recent pyroclastic deposits \citep{miller1985holocene,sieh1986most, Bur1989,bursik1993subplinian,nawotniak2010subplinian, bursik2013digital,bursik2014deposits}, but direct products during the late Pleistocene, tephras preserved within the WCF, had not been analyzed systematically. In this work, we present the detailed stratigraphy and sedimentology of the tephras in the WCF at twelve outcrops near the shoreline of Mono Lake, and highlight their distinct physical properties. Sub-unit correlation has been established for certain tephra units, which enables further investigation and interpretation. Highlights of our work include:

\begin{itemize}
	\item  Each of the rhyolitic tephra units within the WCF is composed of pyroclastic deposits from several eruption pulses. This pulsatory behavior of the Mono Craters is consistent from the late Pleistocene to the Holocene.

	\item The presence of couplets, namely pairs of coarse and fine layers, is a common feature for the tephras in the WCF. Deposition under water, possibly coupled with waning of the eruption through time, is responsible for the development of two separate layers with a sharp internal contact.

	\item Most pumices within the WCF tephras are sub-angular to angular, and are denser than water, with the exception of pumices in B7 and A3, which suggests a distinct degassing history for these layers. The mixing of low-density, rounded lapilli-sized pumice with silty to fine-sandy ash within the uppermost sub-unit of B7 indicates that primary deposits might be transported and deposited elsewhere. For other tephras, it is stressed that the absence of low-density pumice in the deposits does not necessarily mean the absence of it during the eruption.

	\item The abundant obsidians and rhyolitic lithics within many tephras suggest that many eruption pulses involved water-magma interaction and fragmentation of pre-existing domes. Ostracods within D17 and A3 indicate that the eruptive products included lake sediment, and by excluding other possible interpretations such as reworking, 	we suggest that their presence can be regarded as evidence for phreatomagmatic eruption.

	\item Pyroclasts within E18 are characterized by high vesicularity with equant bubble shape, which is unique in the WCF tephras. This is consistent with a different source region, i.e., Mammoth Mountain, for E18 \citep{M14,marcaida2015resolving}. 
	
	\item The Mono Craters were most active during the eruptions of Sequences C and A. Greater thickness and grain size measurement and sub-unit correlation at SEB and TS suggests that C15-a and C11 are characterized by notably greater column height and volume. This implies that C11 may represent deposits from Plinian eruptions instead of a subplinian eruption that is more common to the most recent eruptions from the Mono Craters \citep{sieh1986most,bursik2014deposits}. Well-established sub-unit correlation of C13, C9, and B7 and their comparable but relatively thinner thickness suggest their greater column heights. These features also imply that their vents are located in the southern Mono Craters, or that the associated eruption pulses were short-lived. The overall thinner thickness of the B6 and B5 tephras and the fact that they cannot be correlated well at SEB and TS suggest that their transportation was subject to wind, implying that the corresponding eruptions are of lower column height and smaller volume, or have vents located in the southern Mono Craters. The latter interpretation is preferred due to the limited number of sub-units observed within B6 and B5.

	\item Sub-unit correlation is possible and established for some tephras in the WCF. A more detailed interpretation of the eruptions that produced B7 and tephras in Sequence A is given based on current outcrops. The number of eruption pulses, dispersal direction, the occurrence of pyroclastic flow and surge, and processes taking place during and after the eruptions, such as vent fragmentation and seismic events, are  summarized for all the WCF tephras in Fig. \ref{summary_wcf}.

	\item Thickness measurements, number of sub-units within each tephra layer, and the established sub-unit correlation imply that the vent location of Sequences C-A tephras (except for the basaltic ash layers) migrated from the north (Sequence C) to the south (Sequence B), and then back to the north (Sequence A) of the Mono Craters. Phreatomagmatism was more common for eruptions associated with Sequences C and A tephras, and coincided with highstands of Lake Russell. This is consistent with our inferred vent locations for Sequences C-A tephras.

	\item The stratigraphy from B7* to B7 at SEB and WS indicates a distinct depositional environment, which is probably related to water-level fluctuation of Lake Russell during that time. The stratigraphy of A1 exposed at MC indicates a lower water level of Lake Russell prior to and during the early phase of the A1 eruption; the lake then rose rapidly to a higher level. These inferences agree with previous geological and geochemical analyses, and could add new constraints on the timing and rate of water-level fluctuation of Lake Russell during the late Pleistocene .
\end{itemize}

The present work establishes a framework describing the stratigraphy and sedimentology of the WCF tephras at different sites. It provides new qualitative and quantitative constraints on the eruptive history of the Mono Craters during the late Pleistocene, and proposes different hypotheses to address the uncertainty in the interpretation. At the same time, the work complements the investigation on lacustrine deposits in the WCF \citep{L68,zimmerman2011high}, and with proper understanding of the detailed observations we have presented, could potentially provide new constraints on the water-level history of Lake Russell during the late Pleistocene. The careful treatment of the WCF tephras highlights the importance of recognizing the presence of sub-units within each WCF tephra layer. Our work is expected to benefit futures studies on the WCF tephras in sample collection and data interpretation with greater precision.

\paragraph{Acknowledgements}. This work was supported by National Science Foundation Hazard SEES grant number 1521855 to G. Valentine, M. Bursik, E.B. Pitman and A.K. Patra.  The opinions expressed herein are those of the authors alone and do not reflect the opinion of the NSF. We thank the reviewers for commenting on an earlier version of this manuscript. N. Fontanella, R. Dinnen, M. LaGamba, D. Hyman, R. Sangani, T. J. Scialo are thanked for their help during fieldwork. A. Bevilacqua is thanked for the discussion on the age of the WCF tephras. R. Till is thanked for his help in age calculation. We thank P. Bush for his tutoring and help in working with the SEM machine. This paper is dedicated to the memory of our colleague and friend Sol\`ene Pouget.

\section*{References}
\bibliography{reference}

\section{Tables}
\begin{table}[H]
		{\caption{Terminology defined for different types of deposits observed in this study.}
			{\includegraphics[width=\textwidth]{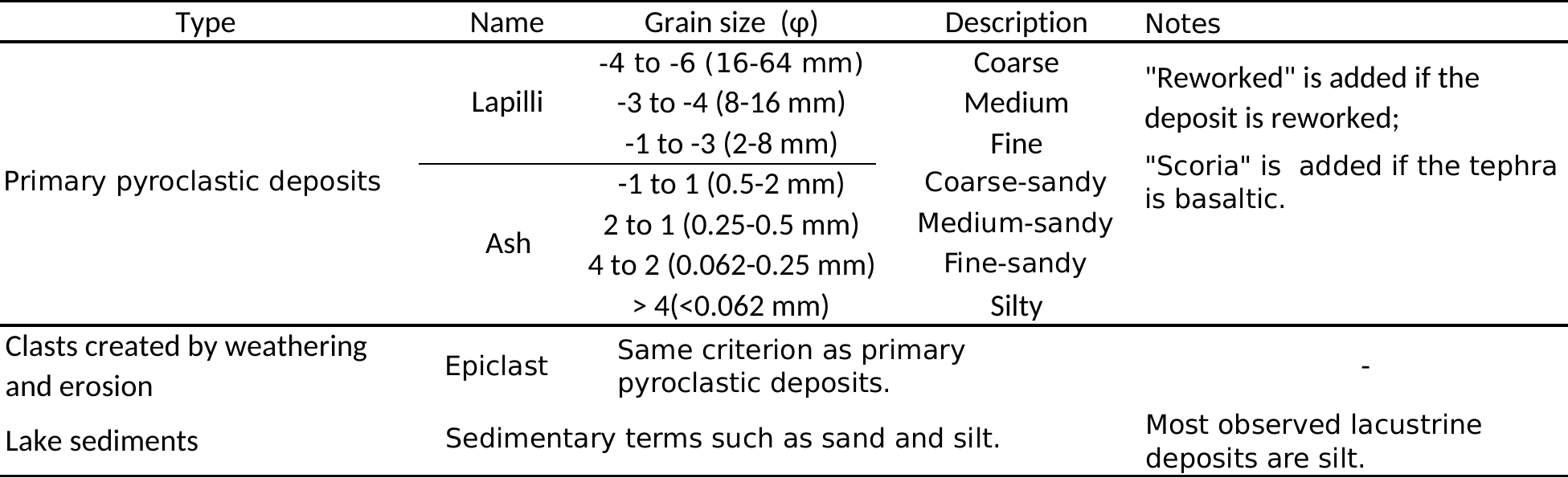}}
			\label{term}}
\end{table}

\begin{table}[H]
	{\caption{Stratigraphic features of selected tephras in Sequences E-C at TS, northwest of Mono Lake. }
		{\includegraphics[width=\textwidth]{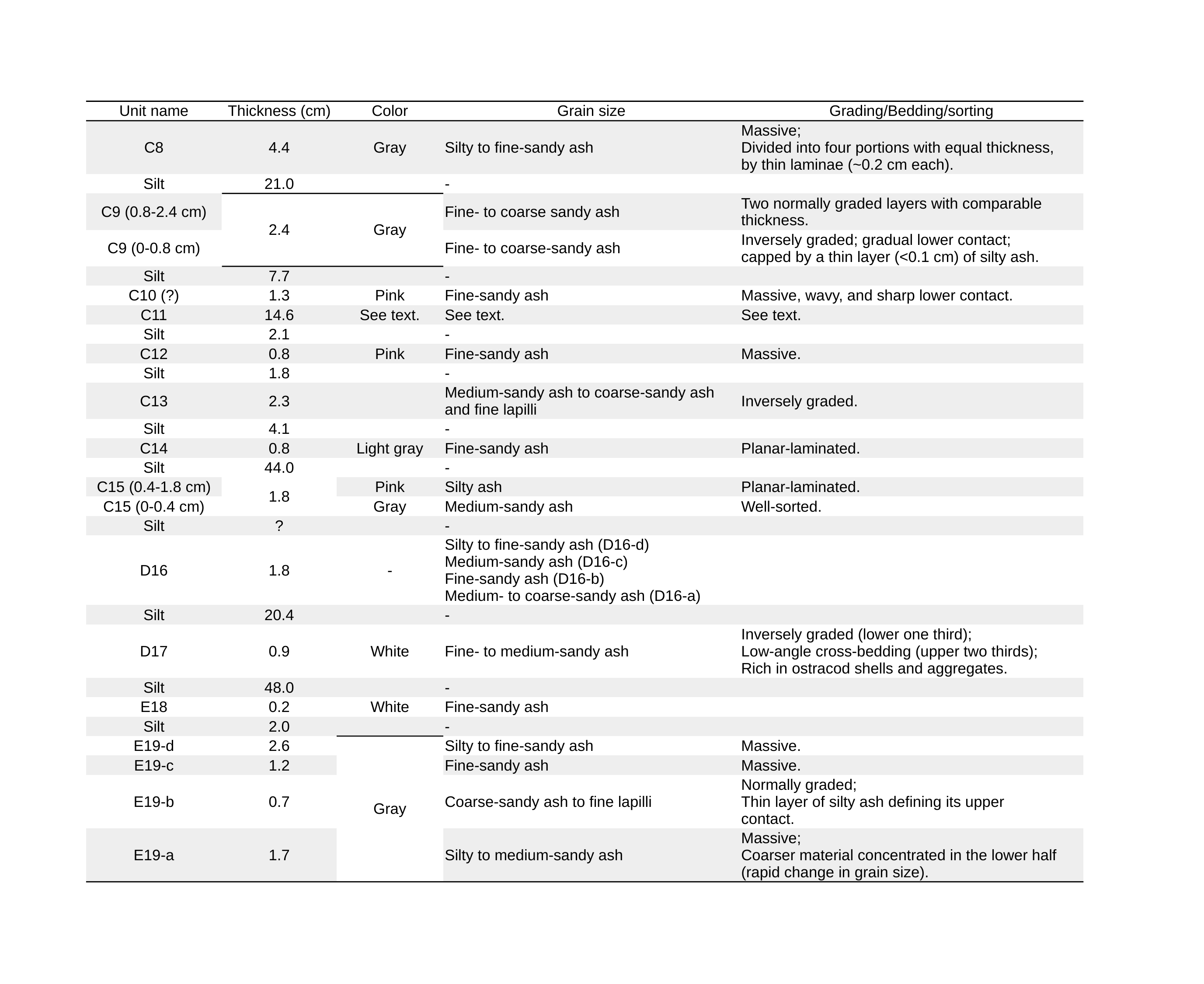}}
		\label{ts_e2c_feature}}
\end{table}

\begin{table}[H]
	{\caption{Stratigraphic features of selected tephras in Sequences D-C at SEB, southeast of Mono Lake.}
		{\includegraphics[width=1.2\textwidth]{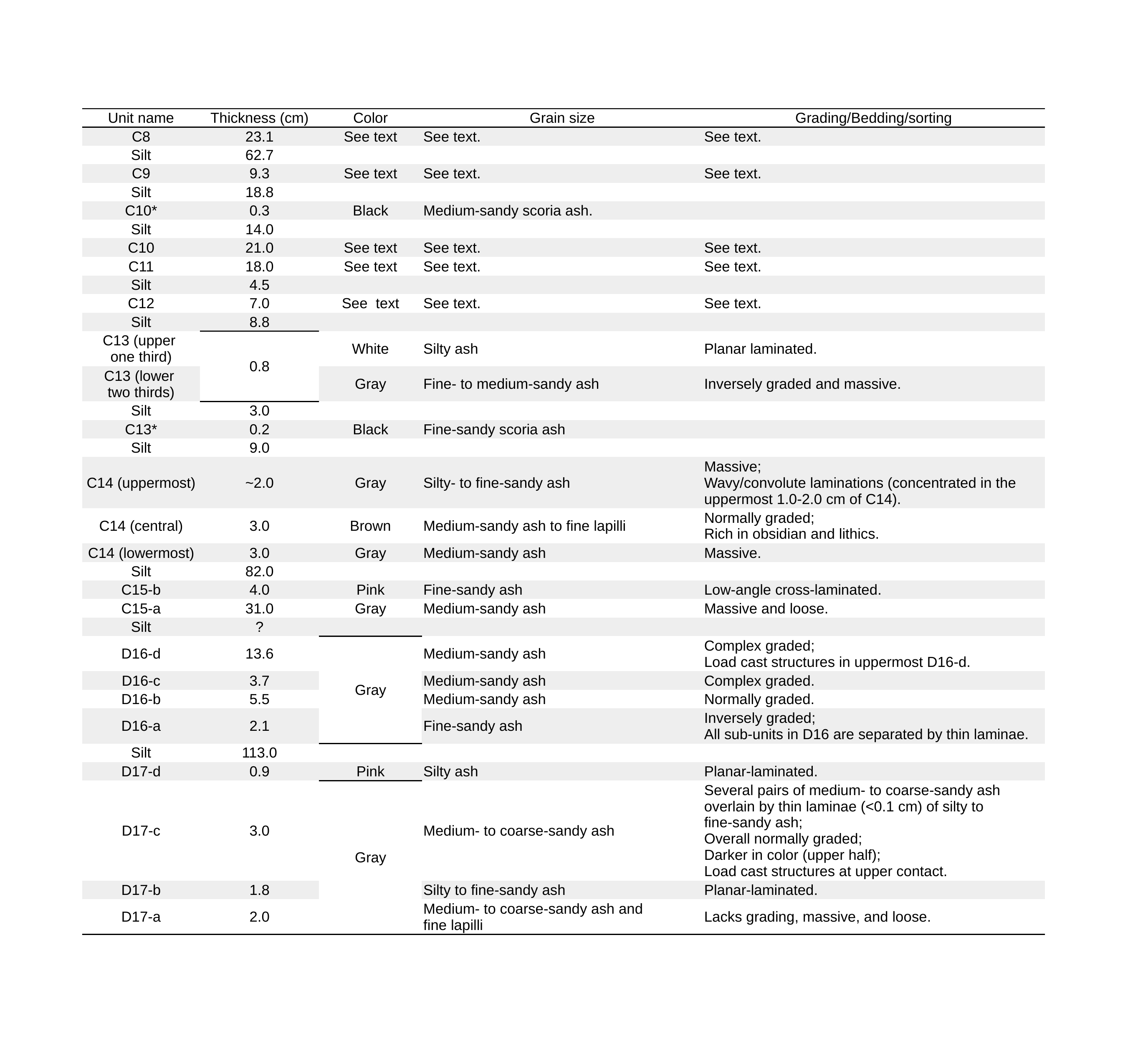}}
		\label{seb_d2c_feature}}
\end{table}

\begin{table}[H]
	{\caption{Stratigraphic features of selected tephras in Sequence B at SEB, TS, WS, and CC.}
		{\includegraphics[width=1.2\textwidth]{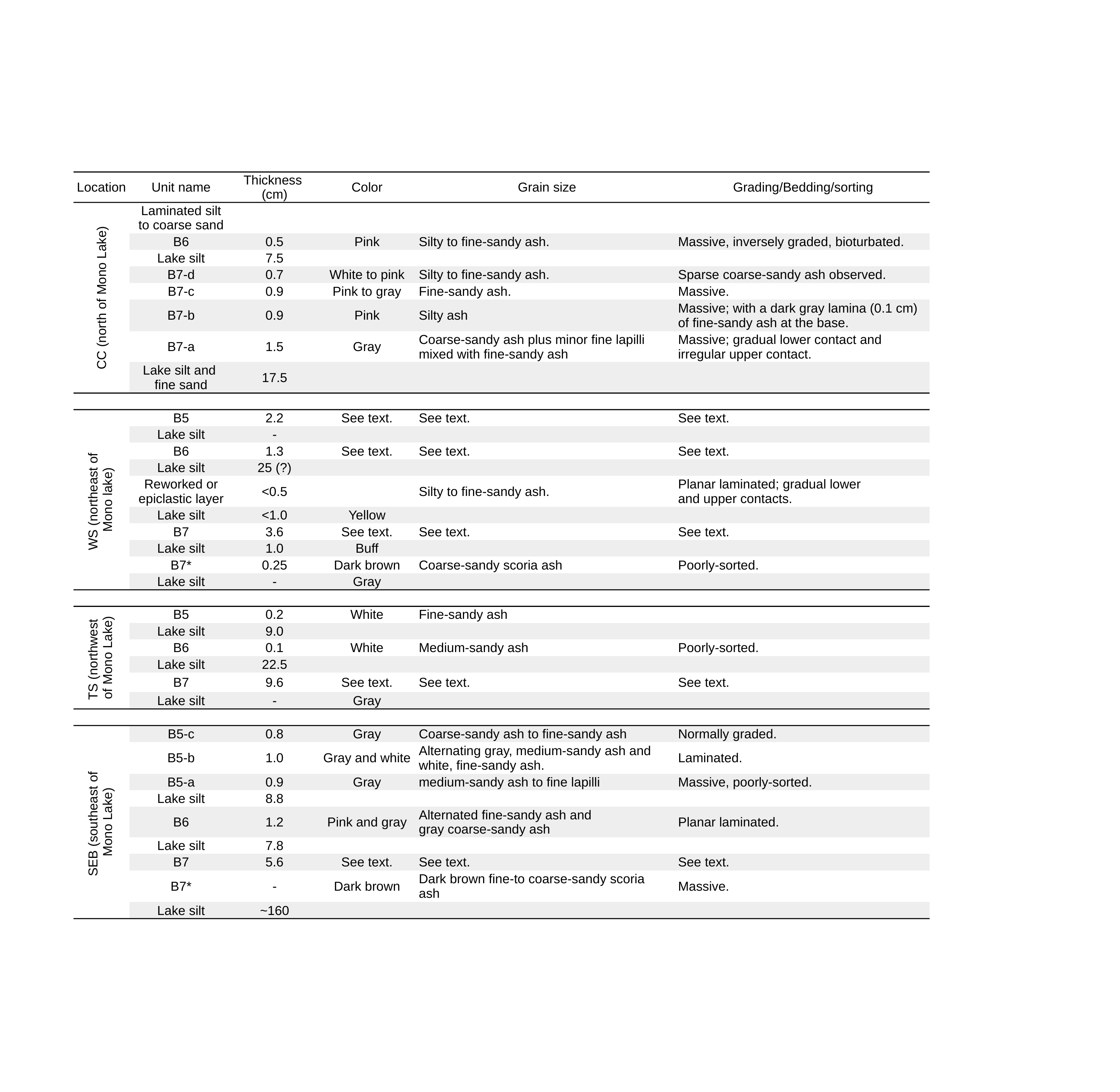}}
		\label{seb_b_feature}}
\end{table}

\begin{table}[H]
	{\caption{Stratigraphic features of A2 at HMC, RC1, RC2, ND1, and WS. Yellow cells represent the correlated sub-unit from the most voluminous eruption pulse of the A2 eruption.}
		{\includegraphics[width=1\textwidth]{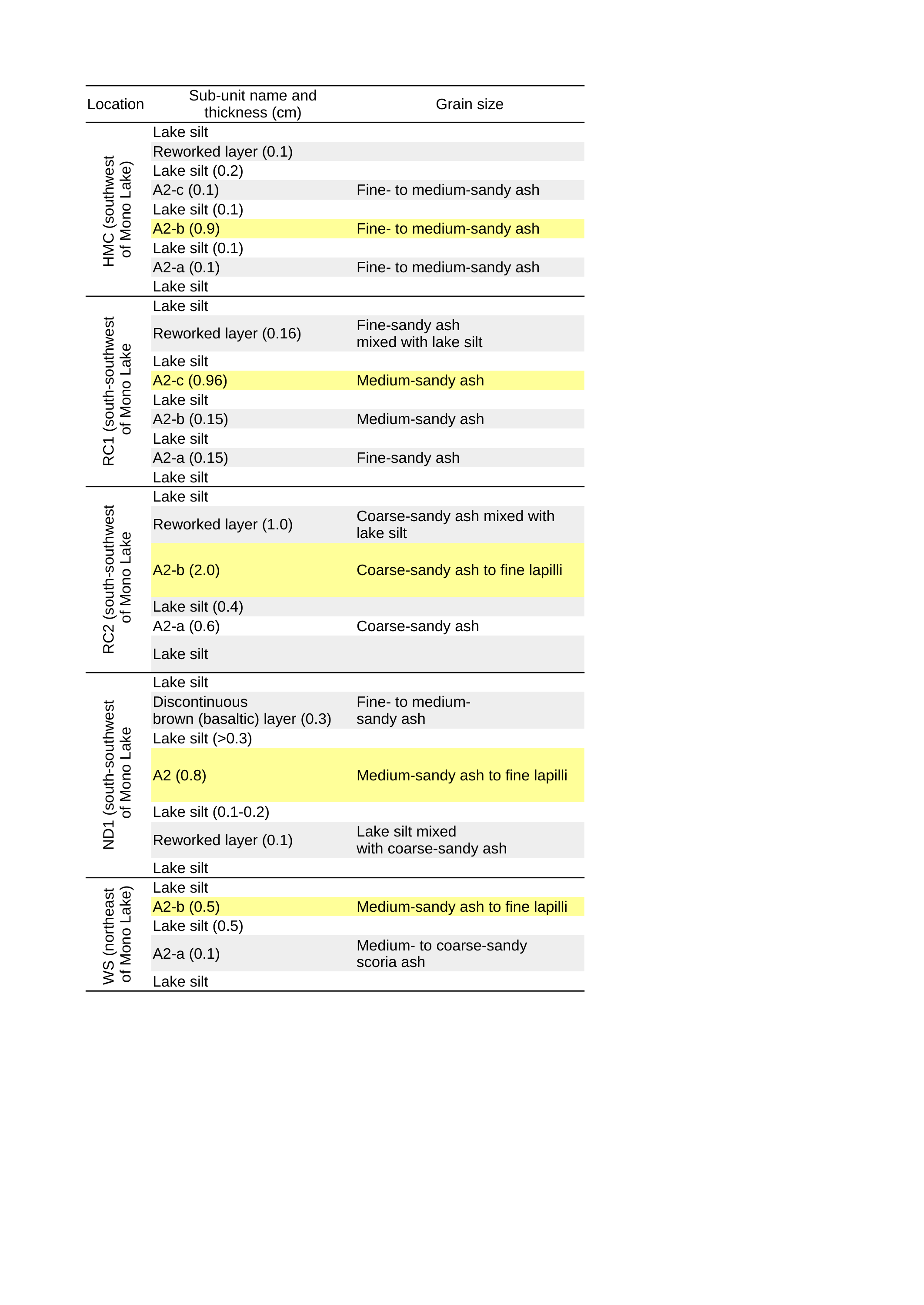}}
		\label{a2_summary}}
\end{table}

\begin{table}[H]
		{\caption{Established sub-unit correlation of tephras in Sequences C-B between sample sites with evidence for correlation.}
			{\includegraphics[width=1.2\textwidth]{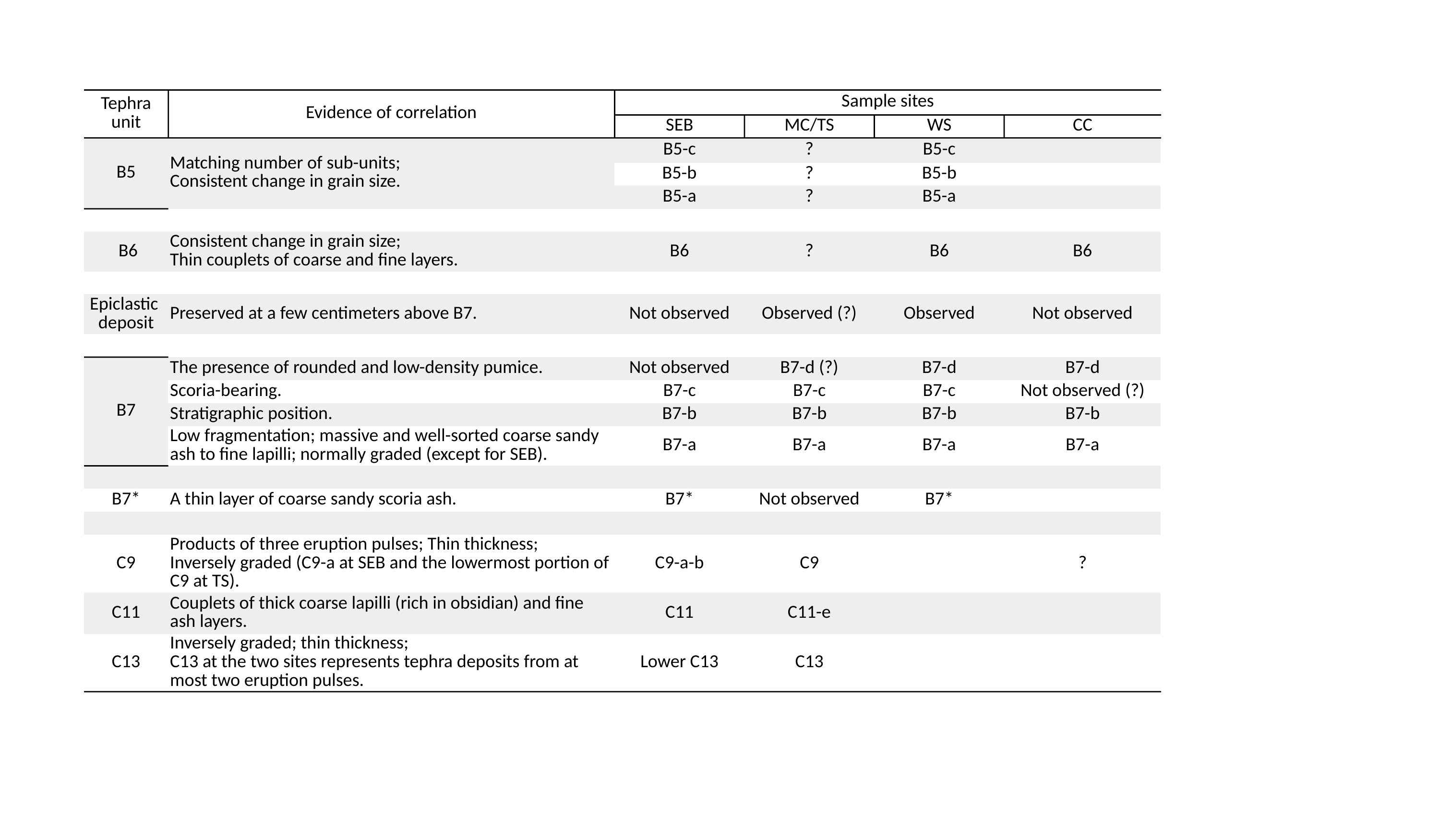}}
			\label{t_cor_b}}
\end{table}

\begin{table}[H]
		{\caption{Established sub-unit correlation of tephras in Sequence A between sample sites with evidence for correlation. Correlations are made separately for sample sites in the east and west of the Mono Craters.}
		{\includegraphics[width=1.2\textwidth]{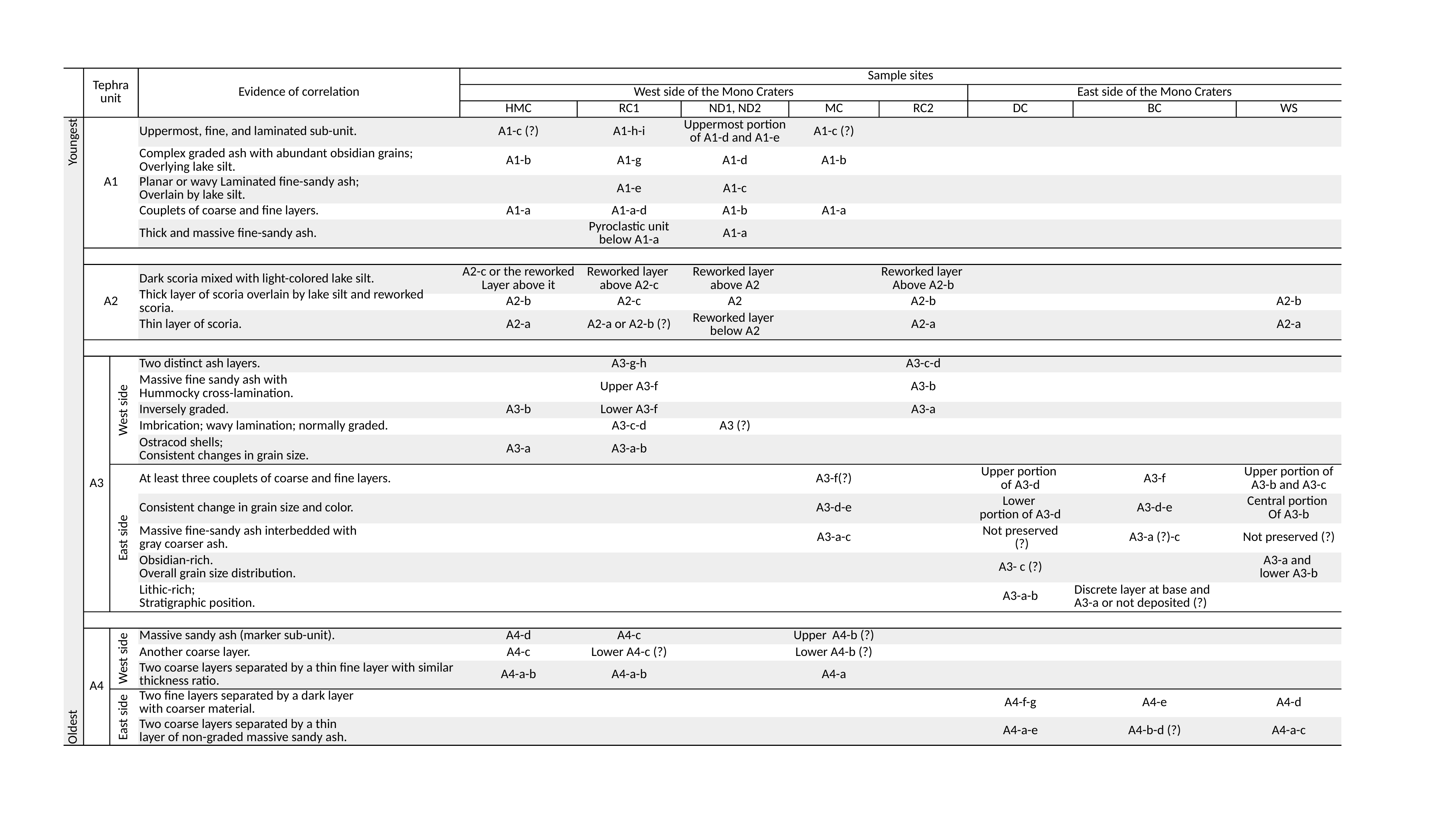}}
		\label{t_cor_a}}
		
\end{table}

\section{Figures}

\end{document}